\documentclass[aps,prd,preprint,nofootinbib,amsmath,amssymb,superscriptaddress]{revtex4-1}
\pdfoutput=1
\usepackage{latexsym}
\usepackage{color}
\usepackage{url}
\usepackage{hyperref}
\usepackage[utf8x]{inputenc} 
\usepackage[usenames,dvipsnames,svgnames,table]{xcolor}
\usepackage{graphicx}
\usepackage{epsfig}  
\usepackage{epsf}   
\usepackage{bm}
\usepackage{subfig}
\usepackage[toc,page]{appendix}
\usepackage{textcomp}
\usepackage{float}
\usepackage{hypcap}
\usepackage[]{hyperref}
\usepackage{booktabs}
\usepackage{multirow}
\usepackage{diagbox}
\usepackage{alphalph}
\usepackage{textcomp}
\usepackage{epstopdf}
\DeclareUnicodeCharacter{2212}{-}
\usepackage{lineno}
\usepackage{soul} 
\usepackage[normalem]{ulem}

\newcommand{\beq}{\begin{equation}}
\newcommand{\eeq}{\end{equation}}
\newcommand{\beqa}{\begin{eqnarray}}
\newcommand{\eeqa}{\end{eqnarray}}
\definecolor{orcidlogocol}{HTML}{A6CE39}

\begin{document}

\title{Extracting the best physics sensitivity from T2HKK: A study on optimal detector volume}

\author{Papia Panda}
\email{ppapia93@gmail.com}
\affiliation{School of Physics, University of Hyderabad, Hyderabad, 500046, India}

\author{Monojit Ghosh}
\email{monojit\_rfp@uohyd.ac.in}
\affiliation{School of Physics, University of Hyderabad, Hyderabad, 500046, India}
\affiliation{Center of Excellence for Advanced Materials and Sensing Devices, Ruder Bo\v{s}kovi\'c Institute, 10000 Zagreb, Croatia}

\author{Priya Mishra}
\email{mishpriya99@gmail.com}
\affiliation{School of Physics, University of Hyderabad, Hyderabad, 500046, India}

\author{Rukmani Mohanta}
\email{rmsp@uohyd.ac.in}
\affiliation{School of Physics, University of Hyderabad, Hyderabad, 500046, India}

\begin{abstract}

T2HK is an upcoming long-baseline experiment in Japan which will have two water Cherenkov detector tanks of 187 kt volume each at distance of 295 km from the source. An alternative project, T2HKK is also under consideration where one of the water tanks will be moved to Korea at a distance of 1100 km. The flux at 295 km will cover the first oscillation maximum and the flux at 1100 km will mainly cover the second oscillation maximum. As physics sensitivity at the dual baseline rely on variation in statistics, dependence of systematic uncertainty, effect of second oscillation maximum and matter density, 187 kt detector volume at 295 km and 187 kt detector volume at 1100 km may not be the optimal configuration of T2HKK. Therefore in this work we have tried to optimize the ratio of the detector volume at both the locations by studying the interplay between the above mentioned parameters. For the analysis of neutrino mass hierarchy, octant of $\theta_{23}$ and CP precision, we have considered two values of $\delta_{\rm CP}$ as $270^\circ$ and $0^\circ$ and for CP violation we have considered the value of $\delta_{\rm CP} = 270^\circ$. These values are motivated by the current best-fit values of this parameter as obtained from the experiments T2K and NO$\nu$A. Interestingly we find that if the systematic uncertainty is negligible then the T2HK setup i.e., when  both the detector tanks are placed at 295 km gives the best results in terms of hierarchy sensitivity at $\delta_{\rm CP} = 270^\circ$, octant sensitivity, CP violation sensitivity and CP precision sensitivity at $\delta_{\rm CP} = 0^\circ$. For current values of systematic errors, we find that neither T2HK, nor T2HKK setup is giving better results for hierarchy, CP violation and CP precision sensitivity. The optimal detector volume which is of the range between 255 kt to 345 kt at 1100 km gives better results in those above mentioned parameters.

\end{abstract}
%======================================================================================

%\arxivnumber{}

%\begin{document}
\maketitle

\section{Introduction}
\label{Introduction}

The phenomenon of neutrino oscillation in which neutrinos change their flavour over a macroscopic time and distance established the fact that neutrinos are massive particles. Mathematically, in the standard three flavour scenario, neutrino oscillation can be described by three mixing angles: $\theta_{13}$, $\theta_{12}$ and $\theta_{23}$, two mass squared differences $\Delta m^2_{21} = m_2^2 - m_1^2$ and $\Delta m^2_{31} = m_3^2 - m_1^2$ with $m_1$, $m_2$ and $m_3$ being the masses of the neutrinos and one Dirac type CP phase $\delta_{\rm CP}$. Among the above mentioned parameters, the parameters $\theta_{12}$, $\theta_{13}$, $\Delta m^2_{21}$ and $|\Delta m^2_{31}|$ have already been measured with very good precision~\cite{Esteban:2020cvm,deSalas:2020pgw,DeSalas:2018rby,Capozzi:2017ipn}. At present, the unknowns are: (i) the sign of $\Delta m^2_{31}$ which can give rise to two possible hierarchy of the neutrino masses i.e.,  $\Delta m^2_{31} > 0$ corresponding to normal hierarchy (NH) and $\Delta m^2_{31} < 0$ corresponding to inverted hierarchy (IH), (ii) the octant of the mixing angle $\theta_{23}$ which can give rise to two possible octants i.e., $\theta_{23} > 45^\circ$ corresponding to higher octant (HO) and $\theta_{23} < 45^\circ$ corresponding to lower octant (LO) and (iii) the unknown phase $\delta_{\rm CP}$. The global analysis of world neutrino data shows a mild preference towards normal hierarchy~\cite{Esteban:2020cvm}. For octant, if the atmospheric data sample is added to the rest of the data sample then the best-fit values of $\theta_{23}$ appears in the lower octant, otherwise the best-fit of $\theta_{23}$ occurs in the higher octant~\cite{Esteban:2020cvm}. Regarding $\delta_{\rm CP}$, currently there is a mismatch in the best-fit values obtained from the ongoing long-baseline experiments T2K~\cite{T2K:2021xwb} and NO$\nu$A~\cite{NOvA:2021nfi}. The best-fit value of $\delta_{\rm CP}$ is around $270^\circ$ for T2K \cite{con_talk_t2k}  and it is around $0^{\circ} / 180^{\circ}$ for NO$\nu$A \cite{con_talk}.

Therefore the aim of next generation experiments will be to establish the nature of the above mentioned unknowns in a firm footing. One of the example of such experiments is T2HK/T2HKK~\cite{Hyper-Kamiokande:2016srs,Hyper-Kamiokande:2018ofw}. T2HK is a proposed long-baseline experiment in Japan. In long-baseline experiments, the accelerated protons hit a target to produce pions and these pions decay to produce an intense beam of muon neutrinos ($\nu_\mu$). These neutrinos are then detected at the far detector. T2HK and T2HKK are the two alternative proposals to detect the neutrinos coming out from the J-PARC accelerator. Under the T2HK project, the proposal is to place two water Cherenkov detector tanks having volume of 187 kt each at a distance of 295 km from the neutrino source. However, the alternative idea is to put one tank at the distance of 295 km in Japan and  the other tank at a distance of 1100 km in Korea. The motivation of this alternative idea is to have excellent CP sensitivity coming from the detector at 295 km due to large statistics and at the same time obtain enhanced hierarchy sensitivity coming from the detector at 1100 km due to large matter effect. In the last few years there are several studies exploring the idea of T2HK and T2HKK in terms of the standard three flavour scenario~\cite{Ghosh:2017ged,Raut:2017dbh,Cho:2019ctv,Agarwalla:2017nld,Fukasawa:2016yue,Chatterjee:2017xkb,Blennow:2014sja,King:2020ydu,Chakraborty:2018dew,Chakraborty:2017ccm,Agarwalla:2017wct,Abe:2017jit,C:2014mmz,Blennow:2020ncm,Ghosh:2019sfi,Coloma:2012wq,Coloma:2012ji,Ballett:2016daj,Evslin:2015pya}. In this paper, we consider the idea of taking a variable detector volume for the T2HKK setup. We define a variable ``$x$" (ranging from 0 kt to 374 kt) which is basically the volume of detector at the baseline of 1100 km having $(374-x)$ kt volume at 295 km. We study the sensitivity of this setup in the standard three flavour scenario as a function of $x$. This is motivated by the following facts. As we increase our baseline from 295 km to 1100 km, the number of events become less and this reduction can decrease the sensitivity. On the other hand, due to the reduction in statistics, systematic uncertainty becomes less dominant on sensitivity which can improve the measurement of oscillation parameters. Further, as the flux corresponding to the 1100 km baseline covers mainly the second oscillation maximum, the effect of matter density becomes less prominent at the 1100 km baseline, so the improvement in the hierarchy sensitivity in this particular case is compromised. However, at the second maximum, the CP sensitivity becomes better than the first oscillation maximum \cite{Hyper-Kamiokande:2016srs}.  Therefore when one moves one detector tank from 295 km to 1100 km, the sensitivity is affected by the interplay of statistics, systematics, second oscillation maximum and matter density. Therefore 187 kt detector volume at 295 km and 187 kt detector volume at 1100 km may not be the optimal configuration for the T2HKK setup. In our study we show that for current estimated systematic uncertainty, higher detector volume at 1100 km is better for hierarchy and $\delta_{\rm CP}$ sensitivity. Interestingly, hierarchy sensitivity at $\delta_{\rm CP} = 270^\circ$, CP violaton sensitivity and CP precision sensitivity at $\delta_{\rm CP} = 0^\circ$ become better when a higher detector volume is placed at 295 km if the estimated systematic uncertainties for this setup becomes negligible. For octant sensitivity higher volume at 295 km is better irrespective of the dependence of systematic errors. In arriving these conclusions, we will point out in detail how the interplay among statistics, systematic errors, second oscillation maximum and matter density affect the optimal detector volume for T2HKK experiment.

The paper is organized as follows. In the next section we will describe the configuration of the T2HK/T2HKK setup which we consider in our calculation. In Section~\ref{res} we will present our numerical results and finally in Section~\ref{sum} we will summarize and conclude.

\section{Simulation Details}
\label{sim}

The experiment T2HK/T2HKK is simulated using the software GLoBES~\cite{Huber:2004ka,Huber:2007ji}. In our calculation we use the configuration as given in \cite{Hyper-Kamiokande:2016srs}. The neutrino source is located at J-PARC having a beam power of $1.3$ MW with a total exposure of $27 \times 10^{21}$ protons on target (POT), corresponding to 10~years of running. We have divided the total run-time into 5 years in neutrino mode and 5 years in anti-neutrino mode (1:1 ratio on $\nu : \bar{\nu}$). To generate the results at 295 km, we have considered a $2.5^\circ$ off-axis flux and for 1100 km, we have considered a $1.5^\circ$ off-axis flux. We have matched the total event rates and event spectrum for both 295 km and 1100 km with respect to the Tables III and IV, and Fig. 13, 14 and 15 of Ref.~\cite{Hyper-Kamiokande:2016srs}. We have considered systematic uncertainties corresponding to overall normalization errors as given in Table VI of Ref.~\cite{Hyper-Kamiokande:2016srs}. We list them here in Table~\ref{table_sys} for reference. 
{In our sensitivity estimation, we have considered two cases of the systematic errors. In one case, all the systematic errors are completely switched off, i.e., there is no effect of systematic errors in the calculation and in the second case, all the systematics are turned on with the values as mentioned in Table~\ref{table_sys}.  In our analysis, the event rate in each bin is altered by the overall normalization factors and they are uncorrelated among different channels.}
For 295 km setup, we have taken 19 bins each for neutrino and antineutrino modes, with neutrino energies $E_{\rm {min}}$ as 0.15 GeV and $E_{\rm {max}}$ as 2.05 GeV, making $100~\rm{MeV}$ energy per bin.  For 1100 km setup, 29 bins (each for neutrino and antineutrino modes) has been taken with $E_{\rm {min}}$ as 0.15 GeV and $E_{\rm {max}}$ as 3.05 GeV, making $100~\rm{MeV}$ energy per bin. We have checked that our results successfully reproduce the physics sensitivities as presented in Ref.~\cite{Hyper-Kamiokande:2016srs}. For our work, we have considered a total detector volume of 374 kt which is distributed in both the locations of 295 km and 1100 km at different ratios. 

\begin{table} 
\centering
\begin{tabular}{|c|c|c|} \hline
Systematics            & 295 km              & 1100 km  \\ \hline
 $\nu_{e}$  & 4.71\%   &     3.84\%  \\ 
$\nu_{\mu}$ & 4.13\%      &  3.83\% \\ 
$\bar{\nu}_e$ & 4.47\%           &    4.11\% \\ 
$\bar{\nu}_\mu$ & 4.15\%         &   3.81\%  \\ 
\hline
\end{tabular}
\caption{The values of systematic errors that we considered in our analysis. The systematic errors are same for signal and background.}
\label{table_sys}
\end{table}  

\begin{table} 
\centering          
\begin{tabular}{|c|c|c|} \hline
Parameters            & True values               & Test value Range  \\ \hline
$\sin^2 \theta_{12}$  & $33.45^\circ$ & Fixed      \\ 
$\sin^2 \theta_{13}$ & $8.62^\circ$                     & Fixed \\ 
$\sin^2 \theta_{23} $ & $42^\circ$/$48^\circ$                  & $39^\circ$ - $51^\circ$\\ 
$\delta_{\rm CP} $       & $ 270^\circ$/$0^\circ$                  & $0^\circ$ - $360^\circ $\\ 
$\Delta m^2_{12}$    & $7.42 \times 10^{-5}~{\rm eV}^2 $ & Fixed \\ 
$\Delta m^2_{31}$    &~ $ 2.51 \times 10 ^{-3}~{\rm  eV}^2$ (NH)~~& $2.43 \times 10 ^{-3}~{\rm  eV}^2$ - $2.60 \times 10 ^{-3}~{\rm  eV}^2$ \\
 \hline
\end{tabular}
\caption{The values of oscillation parameters that we considered in our analysis.}
\label{table_param}
\end{table}

\section{Results}
\label{res}

For the estimation of the sensitivity, we use the Poisson log-likelihood:
\begin{equation}
 \chi^2_{{\rm stat}} = 2 \sum_{i=1}^n \bigg[ N^{{\rm test}}_i - N^{{\rm true}}_i - N^{{\rm true}}_i \log\bigg(\frac{N^{{\rm test}}_i}{N^{{\rm true}}_i}\bigg) \bigg]\,,
 \label{eq:chi2}
\end{equation}
where $N^{{\rm test}}$ is the number of events expected for the values of the oscillation parameters tested for, $N^{{\rm true}}$ is the number of events  expected for the parameter values assumed to be realized in Nature (Asimov dataset) and $i$ is the number of energy bins. The values of the oscillation parameters are taken from Nufit v5.1 \cite{Esteban:2020cvm} and is given in Table \ref{table_param}. In Table \ref{table_param}, the rightmost column gives the $3 \sigma$ ranges of oscillation parameters $\sin^2 \theta_{23}, \delta_{\rm {CP}}$ and $\Delta m_{31}^2$ which we have minimized in the test spectrum, keeping other oscillation parameters fixed.  We will present all our results for the normal hierarchy only.

\subsection{Hierarchy Sensitivity}

\begin{figure}[htpb]
\begin{center}
\hspace{-0.4cm}
\includegraphics[height=55mm,width=75mm]{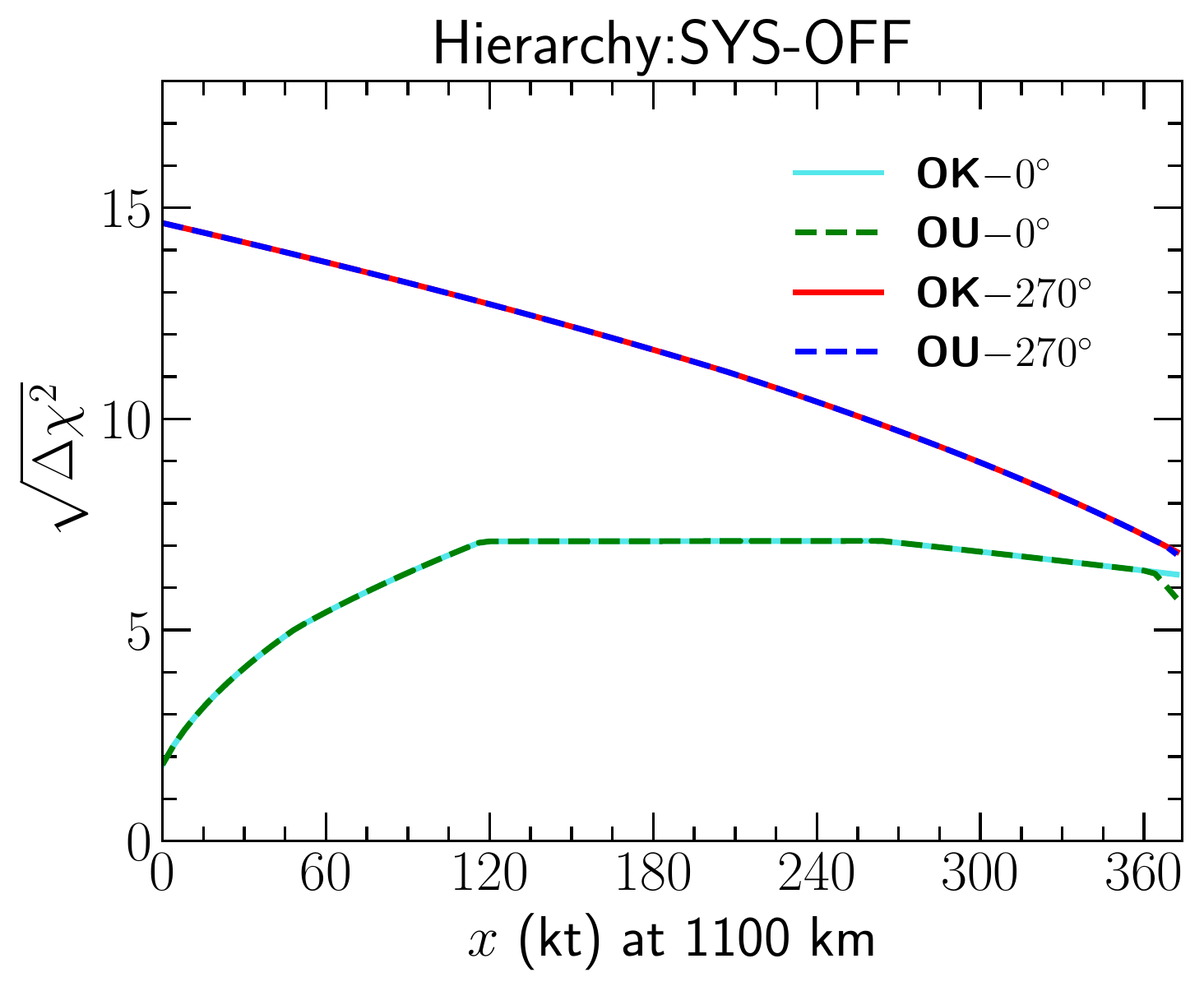}
%\label{hier_1100(a)}
\hspace*{0.1 true cm}
\includegraphics[height=55mm,width=75mm]{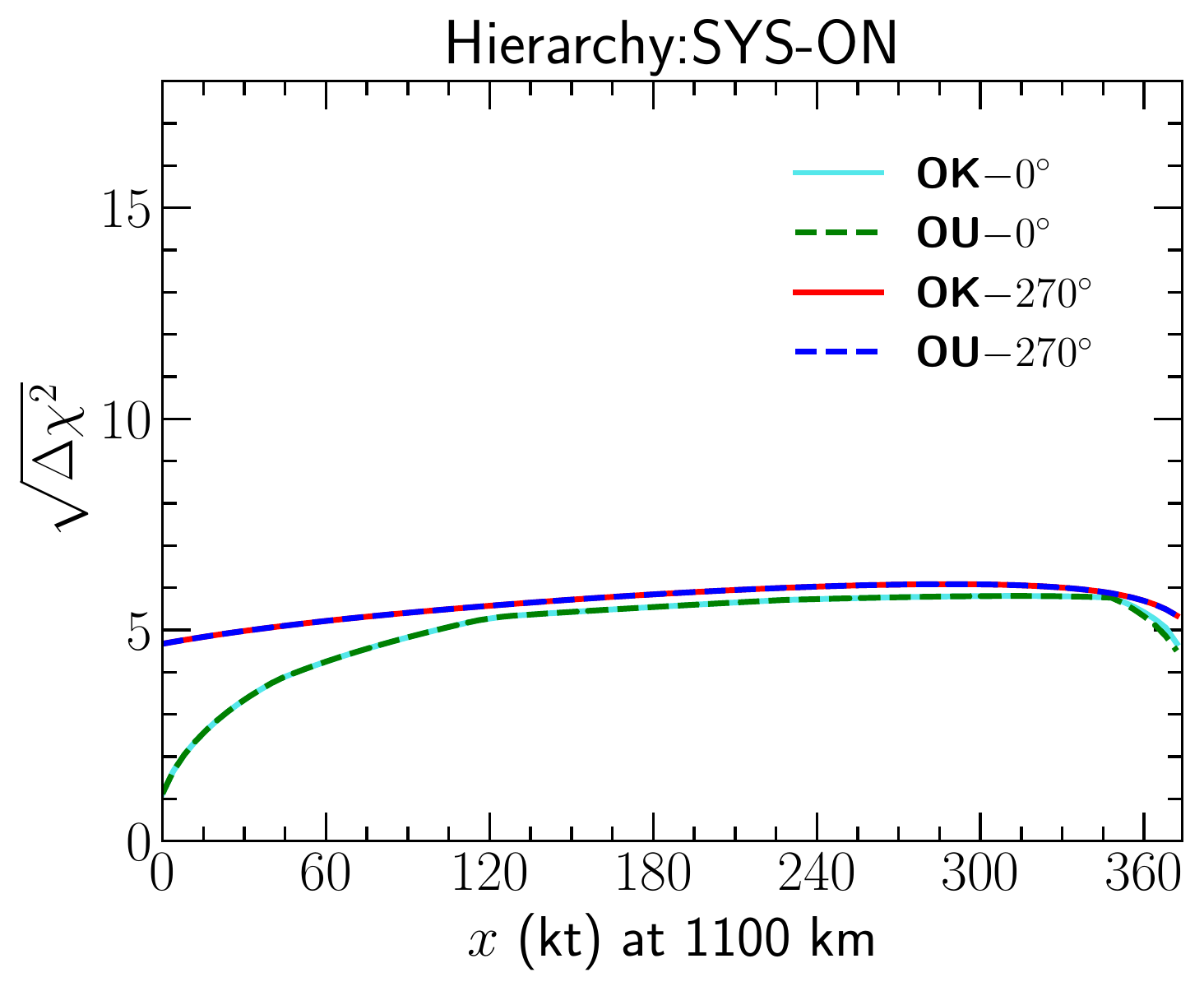}
\caption{Hierarchy sensitivity as a function of $x$. Here $x$ is the detector volume at 1100 km. The left [right] panel corresponding to the case when systematic errors are switched off (SYS-OFF) [on] (SYS-ON). Here `OK' and `OU' refer to octant known and octant unknown respectively. True value of $\theta_{23}$ is taken as $42^\circ$.}
\label{hier_1100}
\end{center}
\end{figure}

Let us first discuss the hierarchy sensitivity of the T2HKK setup by considering a variable detector volume. Hierarchy sensitivity of an experiment is defined by its capability to exclude the wrong hierarchy. In Fig.~\ref{hier_1100}, we have plotted the hierarchy sensitivity as function of $x$, where $x$ is the detector volume at the 1100 km baseline.
The left panel corresponds to the case when systematic errors are switched off whereas the right panel corresponds to the case when the systematic errors are turned on as per given in Table \ref{table_sys}. These panels are generated for true $\theta_{23} = 42^\circ$, which is the current best-fit value of the parameter according to Nufit 5.1. For $\delta_{\rm CP}$, we have considered the true values as $270^\circ$ and $0^\circ$, motivated by the current best-fit of these parameters as obtained from T2K and NO$\nu$A respectively. For each value of $\delta_{\rm CP}$, we have considered two cases: one for octant known referred as `OK' and another for octant unknown referred as `OU'. We have done this to see the effect of octant degeneracy in the hierarchy measurement~\cite{Barger:2001yr,Prakash:2012az,Ghosh:2015ena,Agarwalla:2013ju}. In these panel, the point $x = 0$ kt corresponds to the T2HK setup, where basically both the two detector tanks are placed at a baseline of 295 km, the point $x = 187$ kt corresponds to the T2HKK setup where one tank is placed at a distance of 295 km and another tank is placed at a distance of 1100 km and the point $x = 374$ kt corresponds to the setup where both the detector tanks are located at a distance of 1100 km. The first thing that we notice from Fig.~\ref{hier_1100} is that the sensitivity in the right panel is lower as compared the left panel as far as the value of $x$ is not very large. This shows the fact when the sensitivity is dominated by the statistics (as $x$ increases the overall number of events of this setup decreases), systematic errors play a huge role in the overall sensitivity. At 295 km, number of events are very large and as systematic error directly depends on number of events, the effect of systematic uncertainty on sensitivity is very large at that baseline, resulting a reduced sensitivity at great extent as compared to the sensitivity in absence of systematic errors. But when we go towards 1100 km, as the number of events decrease, the effect of systematic errors on sensitivity becomes less, resulting a small amount of decrease in sensitivity.

Further, from the Fig.~\ref{hier_1100}, we see that as $x$ increases from 0 kt, the hierarchy sensitivity corresponding to $\delta_{\rm CP} = 270^\circ$ decreases and the sensitivity corresponding to $\delta_{\rm{CP}} = 0^\circ$ increases when the systematic errors are switched off. However, when we switch on the systematic errors with the current values, sensitivity for both the curves increases when $x$ increases from 0 kt. This shows a striking feature which was not pointed out earlier that if the systematic errors of T2HK are negligible, then moving a part of the detector to longer baseline does not increase the sensitivity if the true value of $\delta_{\rm CP}$ is $270^\circ$. However, if the true value of $\delta_{\rm CP}$ is $0^\circ$ then moving a part of the detector to a longer baseline can improve the sensitivity irrespective of the value the systematic uncertainty. In these panels we see that for each value of $\delta_{\rm CP}$, octant known and octant unknown curves exhibit same sensitivity implying the octant degeneracy does not affect the hierarchy sensitivity in this setup. From this figure we conclude that if the systematic errors are reduced to very small number (negligible) then T2HK gives best hierarchy sensitivity for $\delta_{\rm CP}$ is $270^\circ$. However, if the systematic errors remain around the current values then the best sensitivity can be obtained if the detector volume is around 255-345 kt (which we call as ``optimized T2HKK") at 1100 km which is different from the standard T2HKK setup.

\begin{figure}[htpb]
    \centering
    \hspace{-0.4cm}
    \includegraphics[height=55mm,width=75mm]{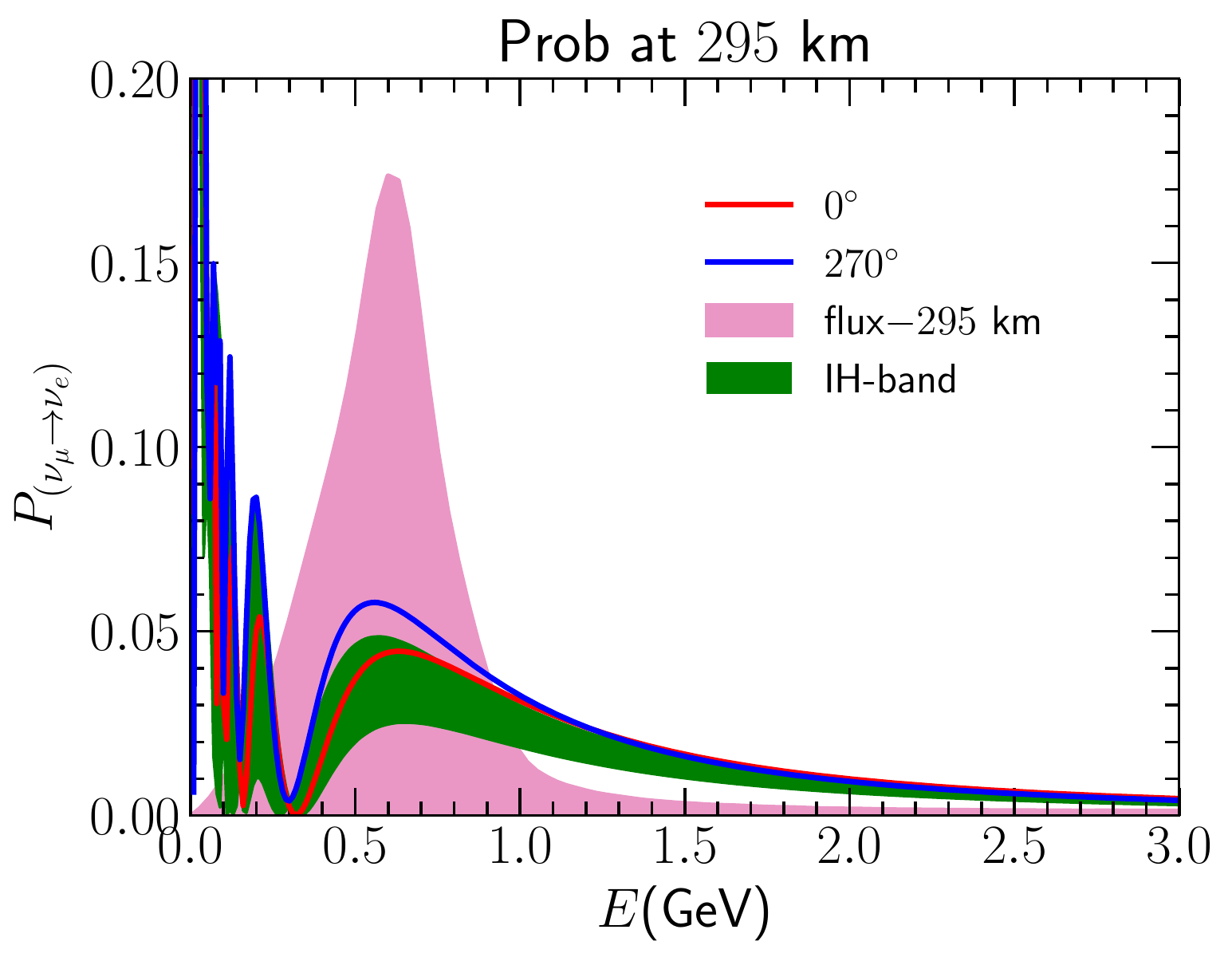}
    \hspace*{0.1 true cm}
    \includegraphics[height=55mm,width=75mm]{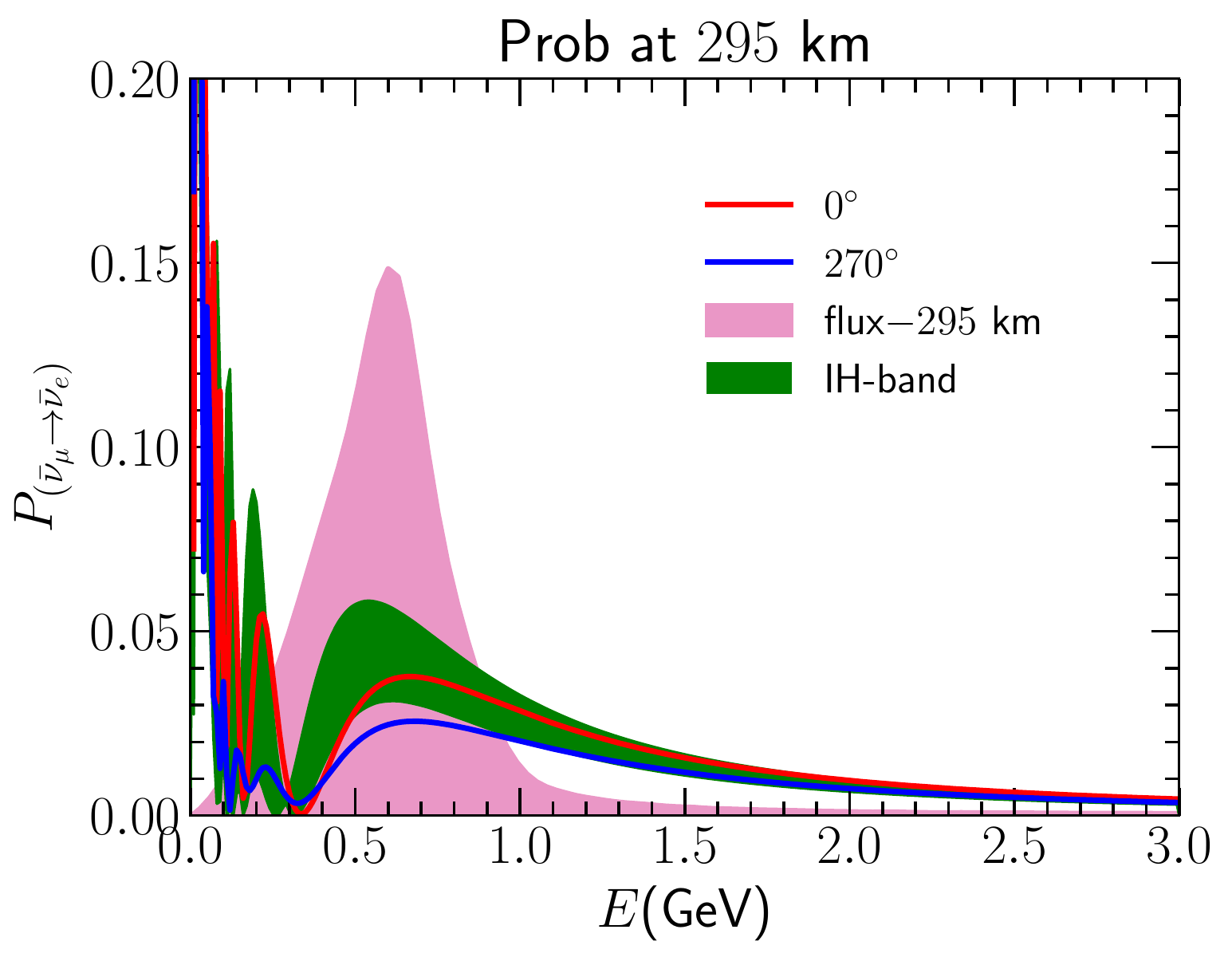} \\
    \hspace{-0.4cm}
    \includegraphics[height=55mm,width=75mm]{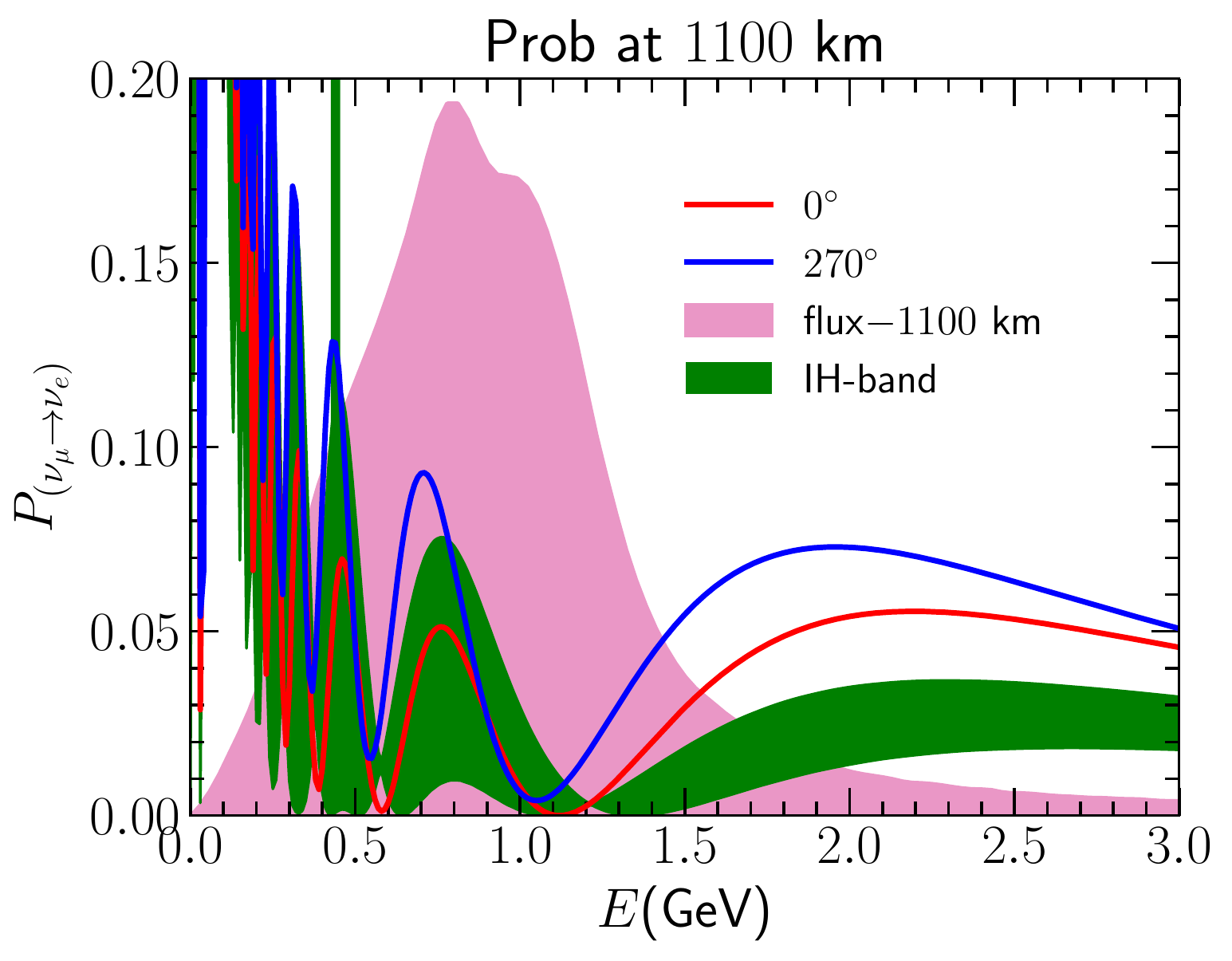}
    \hspace*{0.1 true cm}
    \includegraphics[height=55mm,width=75mm]{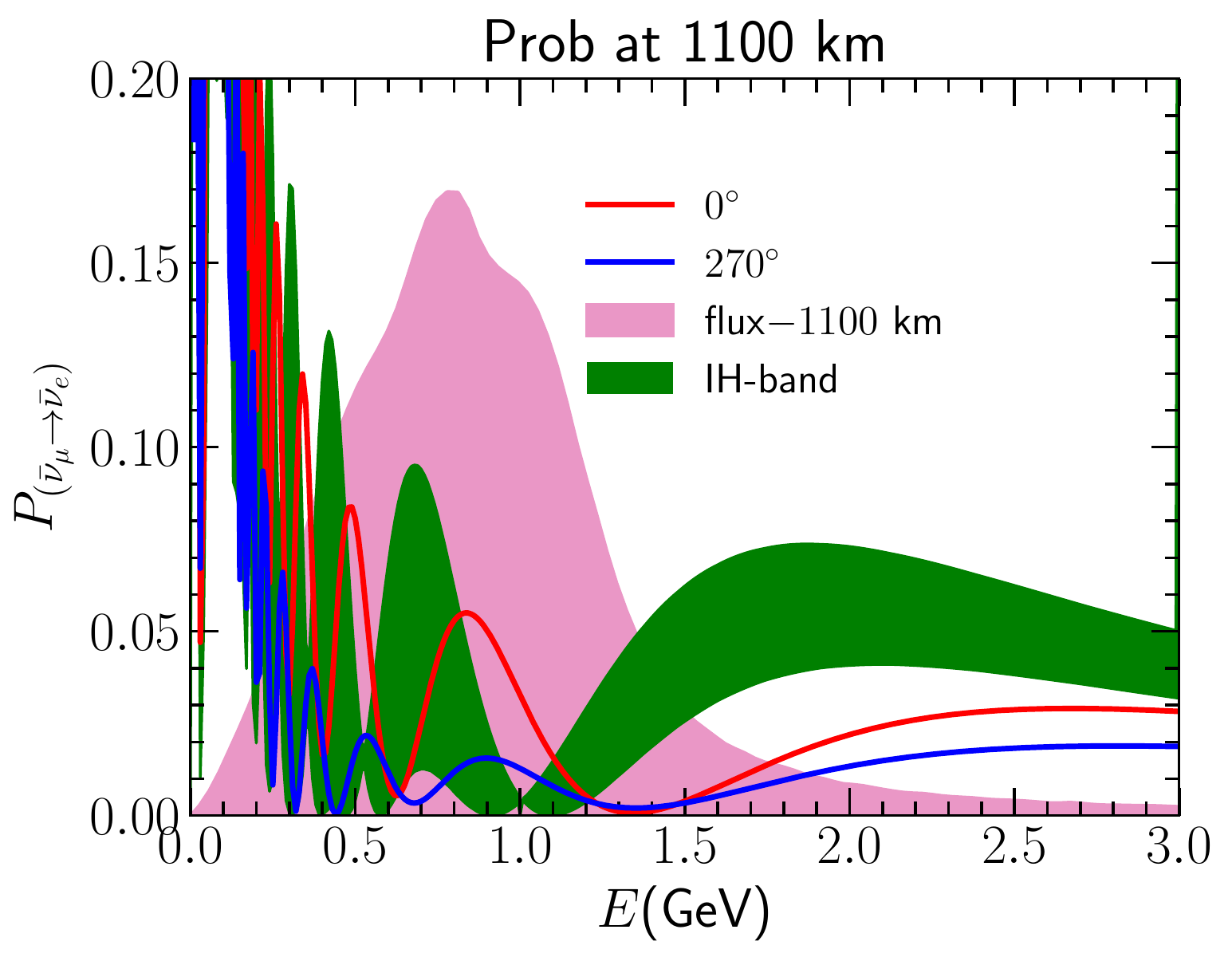}
    \caption{Upper (lower) row shows oscillation probabilities at 295 km (1100 km). Left (right) column depicts the probabilities for (anti) neutrino mode. In each plot, pink shaded area represents corresponding fluxes, green shaded region is for inverted hierarchy band and red (blue) solid line shows the probability for $\delta_{\rm CP} = 0^{\circ} (270^{\circ})$ in normal hierarchy.}
    \label{prob_hier}
\end{figure}

To understand the underlying physics of Fig.~\ref{hier_1100}, in Fig.~\ref{prob_hier}, we have plotted the appearance channel probability ($P_{\nu_{\mu} \rightarrow \nu_e}$) as a function of energy. For long-baseline experiments, the hierarchy sensitivity comes from the appearance channel~\cite{Akhmedov:2004ny}.
The top row is for 295 km and the bottom row is for 1100 km. These panels are generated for $\theta_{23} = 42^\circ$. In each row the left panel is for neutrinos and the right panel is for antineutrinos. In each panel, the pink shaded region corresponds to the flux at that baseline and polarity. The area covered by the shaded region reflects the energy region to which the experiment is sensitive to. In these panels, the red and blue curves are for normal hierarchy for two different values of $\delta_{\rm CP} = 0^\circ$ and $270^\circ$ respectively and the green band is for inverted hierarchy. The width of the green band is due to the variation of $\delta_{\rm CP}$ in its full range. Therefore, the separation between the red curve and the green band is proportional to the hierarchy sensitivity at $\delta_{\rm CP} = 0^\circ$ and the separation between blue curve and green band is proportional to the hierarchy sensitivity at $\delta_{\rm CP} = 270^\circ$. From the panels it clear that the baseline 295 km mainly covers the first oscillation maximum whereas the baseline 1100 km covers mainly the second oscillation maximum. 
\begin{itemize}
    \item First let us discuss the case for $\delta_{\rm CP} = 270^\circ$. From the panels we note that the separation between the blue curve and green band is much higher for the baseline of 295 km as compared to 1100 km in the energy region that is covered by the corresponding fluxes. This is the reason, when systematic errors are off, increasing the value of $x$ from 0 kt causes a reduction in the sensitivity (cf. Fig.~\ref{hier_1100}). However, if the flux at 1100 km would have covered the first oscillation maximum, then the separation between the blue curve and green band would have been much higher for 1100 km as compared to 295 km. Note that the separation of the oscillation probabilities in normal hierarchy and inverted hierarchy depend on the matter effect. For a given baseline, the matter effect is proportional to the energy of the neutrinos. The first maximum occurs at a higher energy as compared to the second maximum, therefore the separation of the curves corresponding to different hierarchies are higher in the first maximum as compared to the second maximum. When the systematic errors are on, we see that the sensitivity increases as we increase $x$. This is because as we increase $x$, the total number of events gets reduced and hence the sensitivity becomes less dominant on the systematic uncertainties which causes the sensitivity to rise as we increase $x$. Or in other words, in presence of systematic errors, the reduction in the sensitivity is higher when $x$ is small and lower when $x$ is large. But here it is important to note that the best sensitivity does not come at $x = 374$ kt. After a certain value of $x$, the sensitivity starts to decrease. This feature we will see in the other figures too. This is because after a certain value of $x$, the statistics reduces so much that even the improvement in effect of systematic errors could not improve the sensitivity. From the above discussion, we understand that the improvement in hierarchy sensitivity at 1100 km is not due to the matter effect, but is totally due to the interplay between statistics and systematic errors.
    
    \item For $\delta_{\rm CP} = 0^\circ$, we see that the red curve overlaps with the green band for 295 km. But for 1100 km, there is some finite separation between the red curve and the green band. This is the reason why the sensitivity increases as we increase $x$ from 0 kt even when the systematic errors are switched off. For the case when the systematic errors are switched on, the reason for the improvement in the sensitivity is two fold: (i) sensitivity coming from the 1100 km and (ii) improvement in the effect of systematic errors due to less statistics. 
\end{itemize}

\begin{figure}[htpb]
\begin{center}
\hspace{-0.4cm}
\includegraphics[height=55mm,width=75mm]{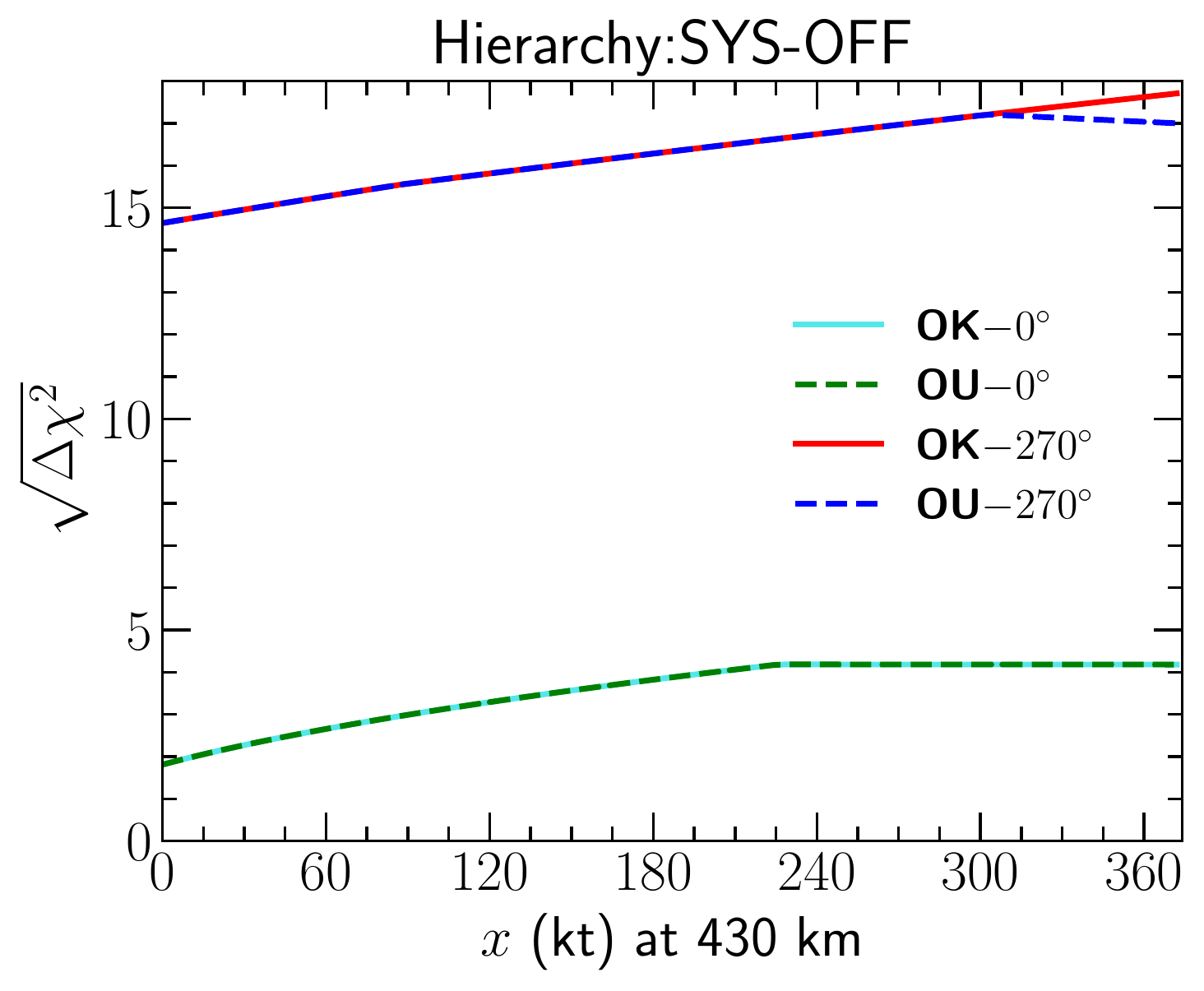}
\hspace*{0.1 true cm}
\includegraphics[height=55mm,width=75mm]{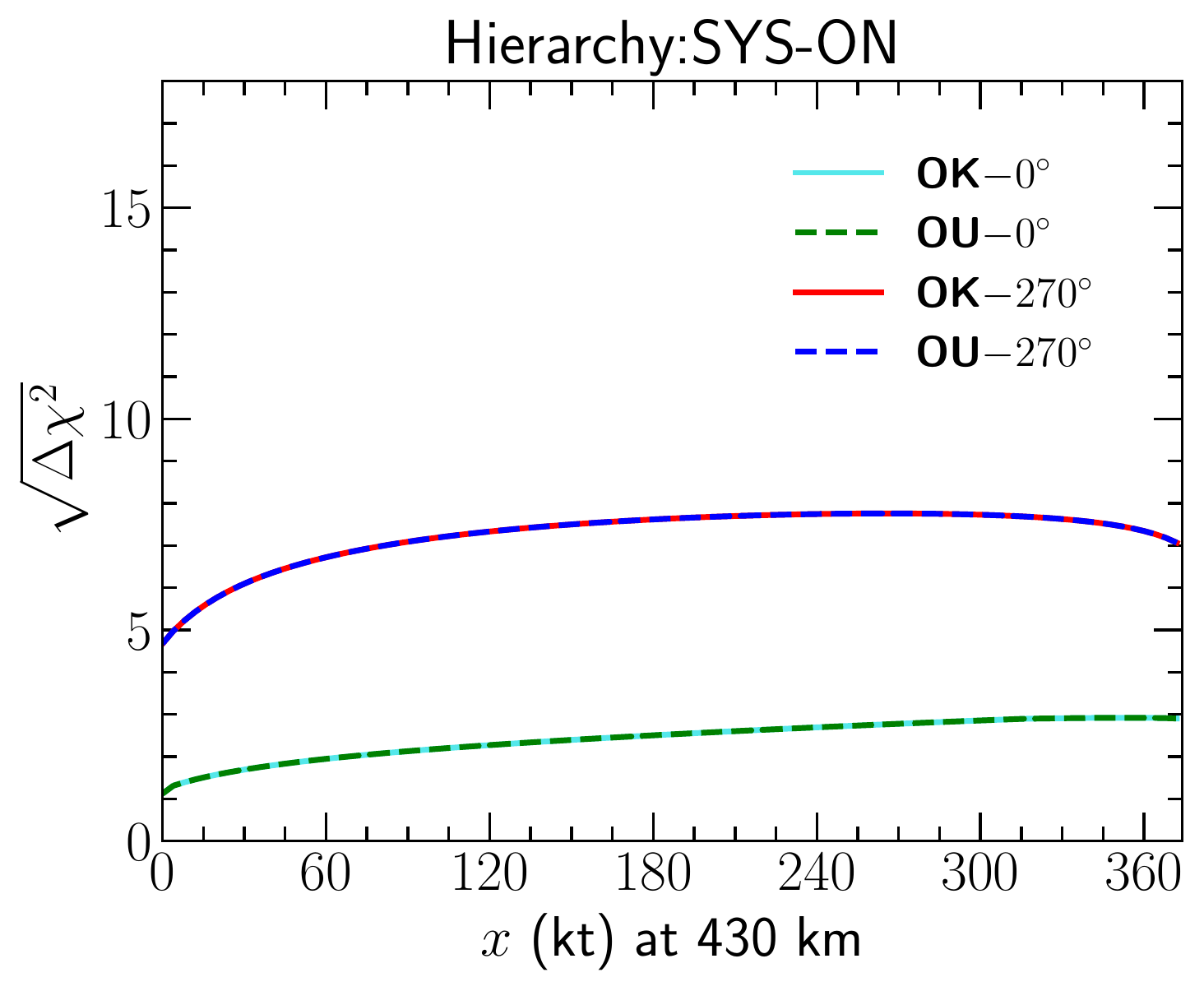}
\caption{Hierarchy sensitivity as a function $x$. Here $x$ is the detector volume at a hypothetical distance of 430 km. The left (right) panel corresponding to the case when systematic errors are switched off (on). Here `OK' and `OU' refer to octant known and octant unknown respectively. True value of $\theta_{23}$ is $42^\circ$.}
\label{hier_430}
\end{center}
\end{figure}

Now we will try to see what would have happened if the flux at 1100 km covers the first oscillation maximum instead of second oscillation maximum. As we do not have the flux at 1100 km which covers the first oscillation maximum, we tried to adjust the baseline such that the existing flux at 1100 km coincides with the first maximum. We find that baseline to be 430 km. In Fig.~\ref{hier_430}, we plot the same as that of Fig.~\ref{hier_1100} except the fact that here we assume that the Korean detector is placed at a hypothetical distance of 430 km.
From this figure we see that the hierarchy sensitivity for $\delta_{\rm CP} = 270^\circ$ increases as $x$ increases from 0 kt when the systematic errors are switched off. This is the stark contrast as compared to the case when the flux cover the second maximum at 1100 km. As in this case the sensitivity also comes from the first maximum, moving the detector volume from 295 km to 430 km increases the hierarchy sensitivity even when the systematic are switched off. Regarding the rest of the curves the behaviour is same as that of Fig.~\ref{hier_1100} and can be explained in the similar fashion. 

\subsection{Octant Sensitivity}

\begin{figure}[htpb]
    \centering
    \hspace{-0.4cm}
    \includegraphics[height=55mm,width=75mm]{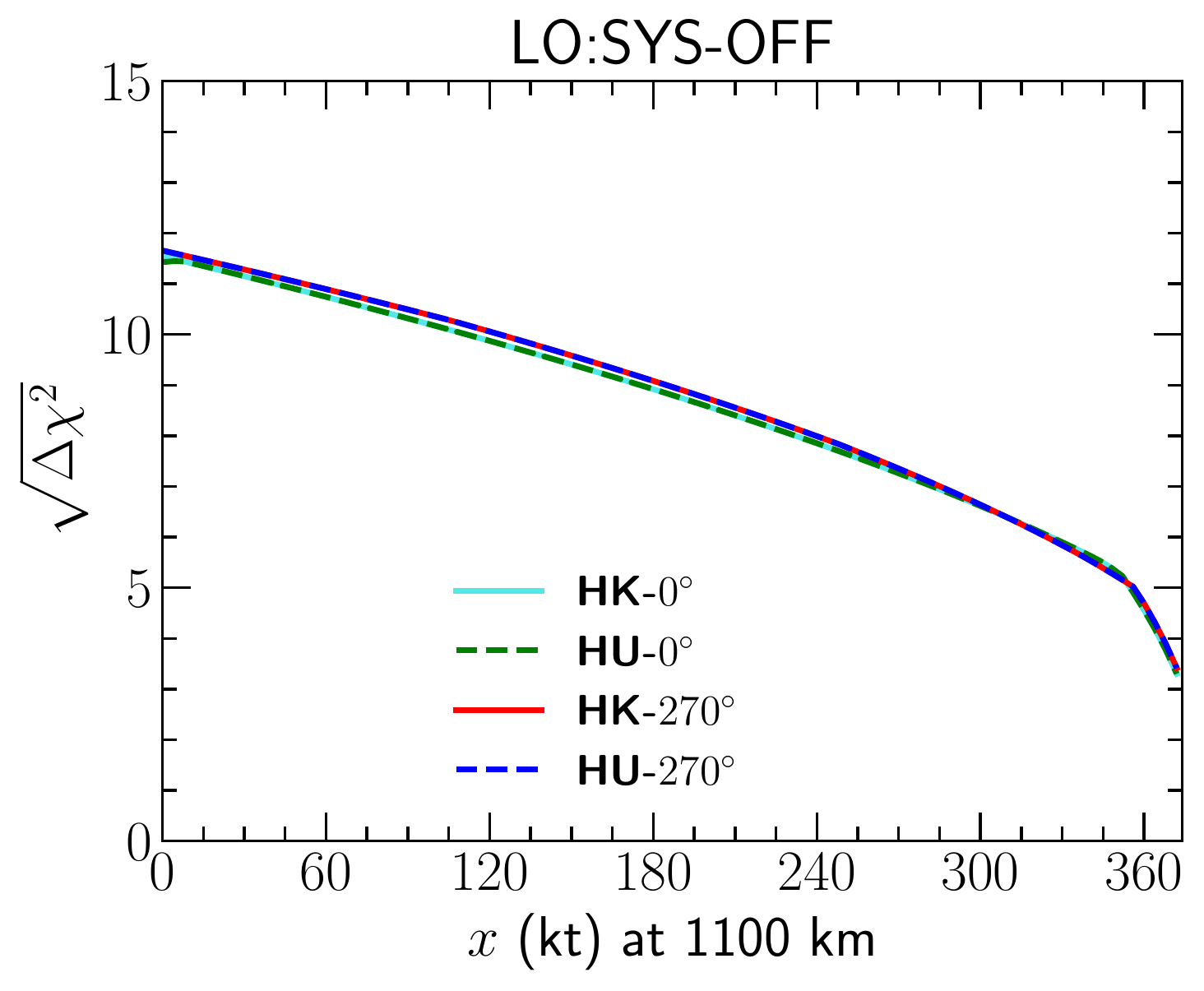}
    \hspace*{0.1 true cm}
    \includegraphics[height=55mm,width=75mm]{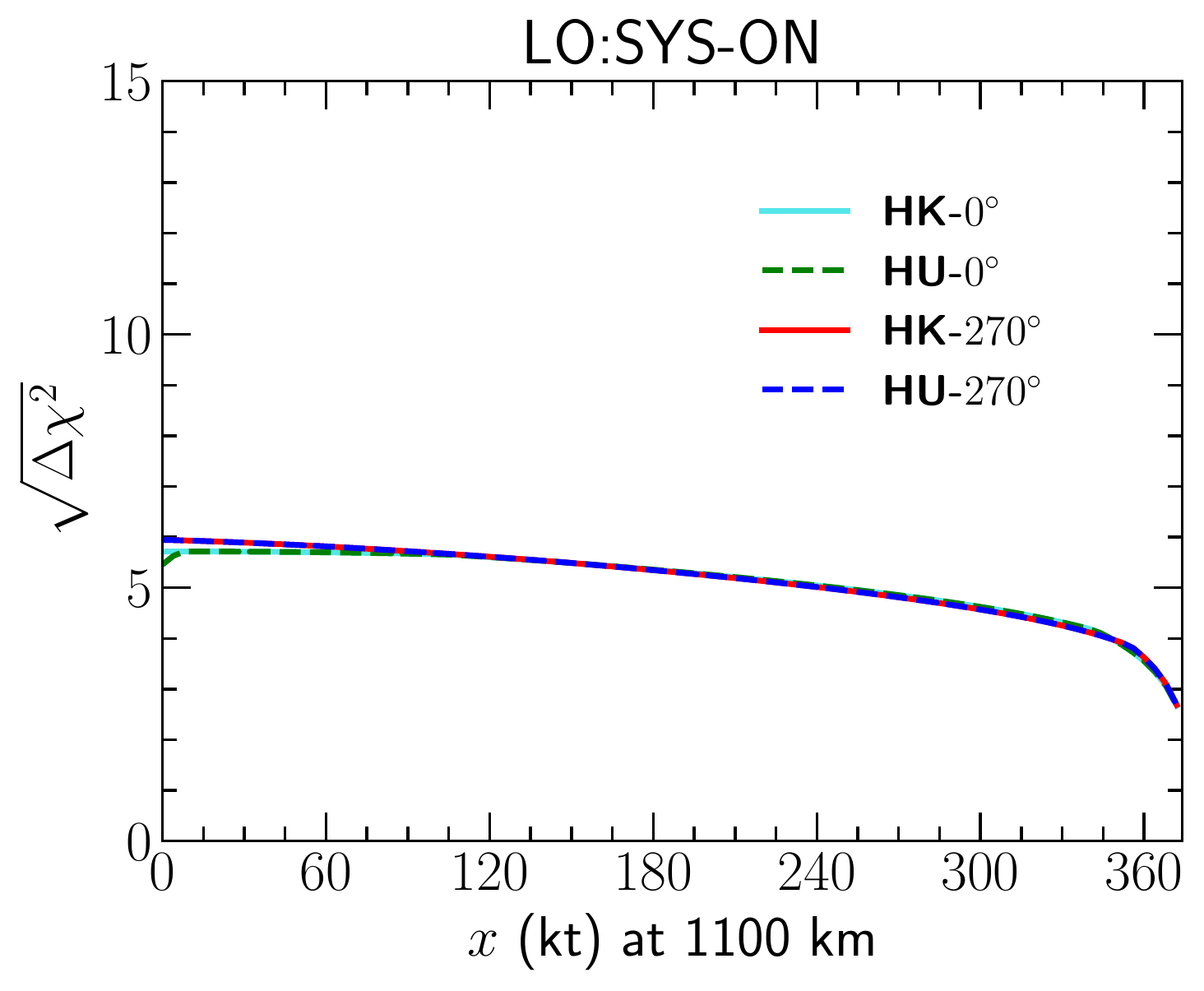}  \\
    \hspace{-0.4cm}
    \includegraphics[height=55mm,width=75mm]{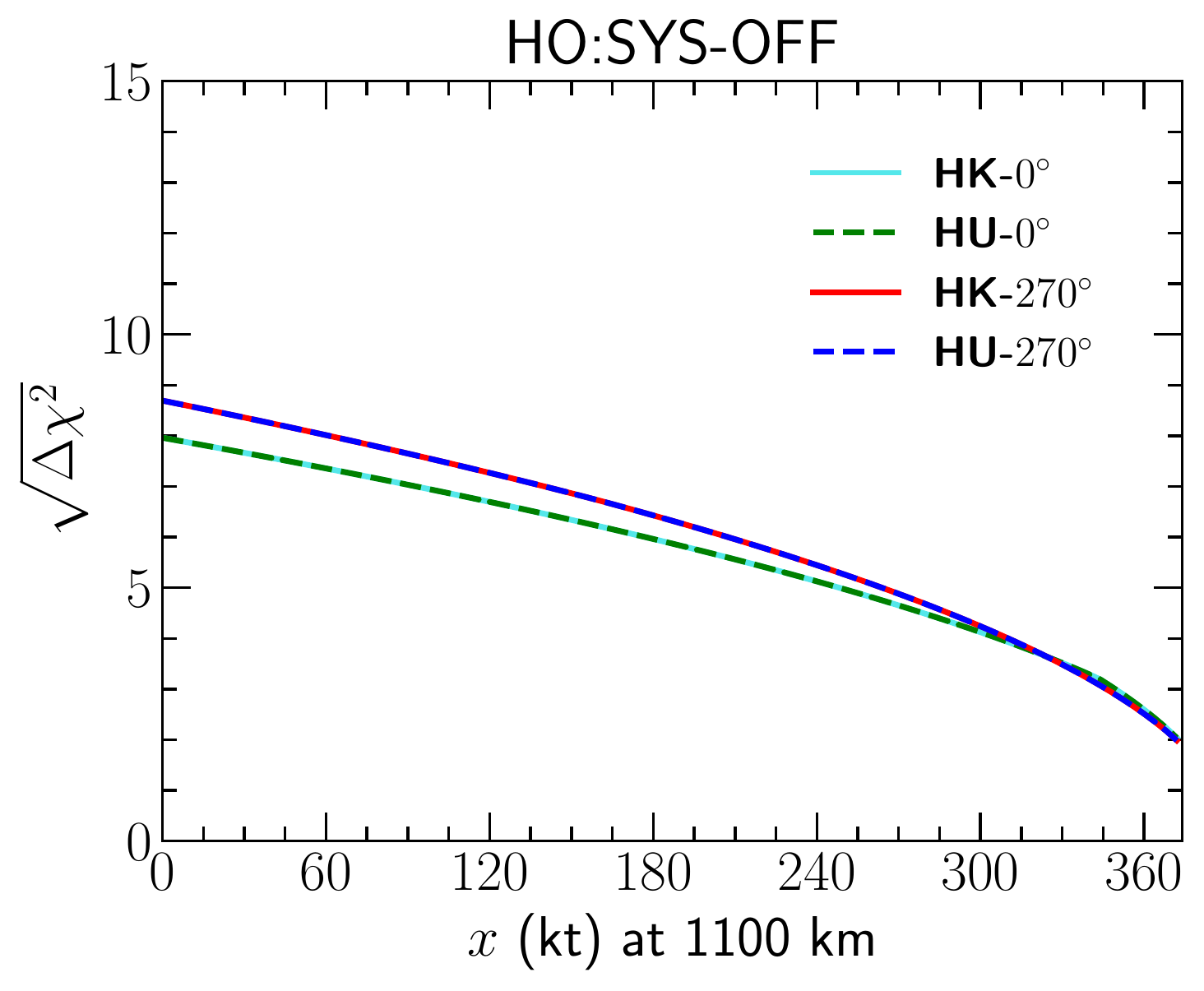}
    \hspace*{0.1 true cm}
    \includegraphics[height=55mm,width=75mm]{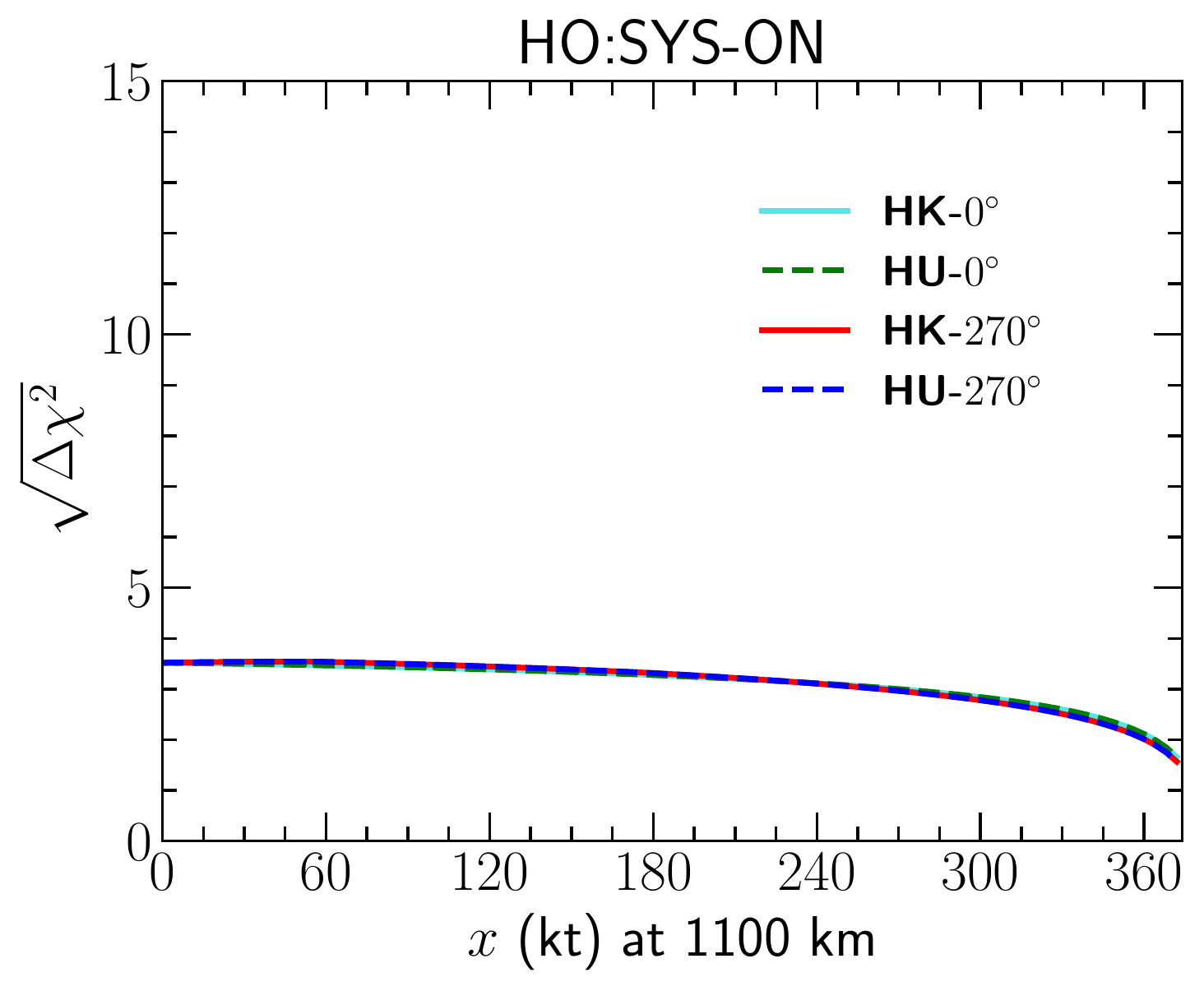}
    \caption{Octant sensitivity as a function of $x$ (detector volume at 1100 km). Upper (lower) row shows octant sensitivity when $\theta_{23}^{\rm {true}}=42^{\circ} (48^{\circ})$. Left (right) column reflects the same without (with) systematic error. In each plot, `HK (HU)$-0^{\circ} [270^{\circ}]$' stands for hierarchy known (unknown) with $\delta_{\rm {CP}}^{\rm {true}} =0^{\circ} [270^{\circ}] $. }
    \label{oct_1100}
\end{figure}

Now let us discuss the octant sensitivity of the T2HKK setup with respect to $x$. In Fig.~\ref{oct_1100}, we have plotted the octant sensitivity as a function of $x$. 
Here $x$ is the detector volume at $1100$ km. In this figure, the left column is for the case when systematic errors are switched off and the right column is for the case when systematic errors are switched on. The top row refers to the true lower octant corresponding to $\theta_{23} = 42^\circ$ and the bottom row refers to the true higher octant corresponding to $\theta_{23} = 48^\circ$. For $\delta_{\rm CP}$, we have considered the true values as $270^\circ$ and $0^\circ$. For each value of $\delta_{\rm CP}$, we have considered two cases: one for hierarchy known referred as `HK' and another for hierarchy unknown referred as `HU'. We have done this to see the effect of hierarchy degeneracy in the octant measurement. From this figure we note that except the case of true higher octant with systematic off, the octant sensitivity corresponding to $\delta_{\rm CP} = 270^\circ$ and $0^\circ$ are same. Further we also see that hierarchy degeneracy does not play any role in the determination of true octant. In all the four panels we see that as $x$ increases from 0 kt, the sensitivity decreases. This is true for both true octants, both the values of $\delta_{CP}$ and irrespective of the fact that systematic errors are turned off or on. Therefore from this figure, we conclude that for octant sensitivity, the best sensitivity comes from the T2HK setup i.e., the setup when both the detectors are placed at a distance of 295 km.  

\begin{figure}[htpb]
    \centering
    \hspace{-0.4cm}
    \includegraphics[height=55mm,width=75mm]{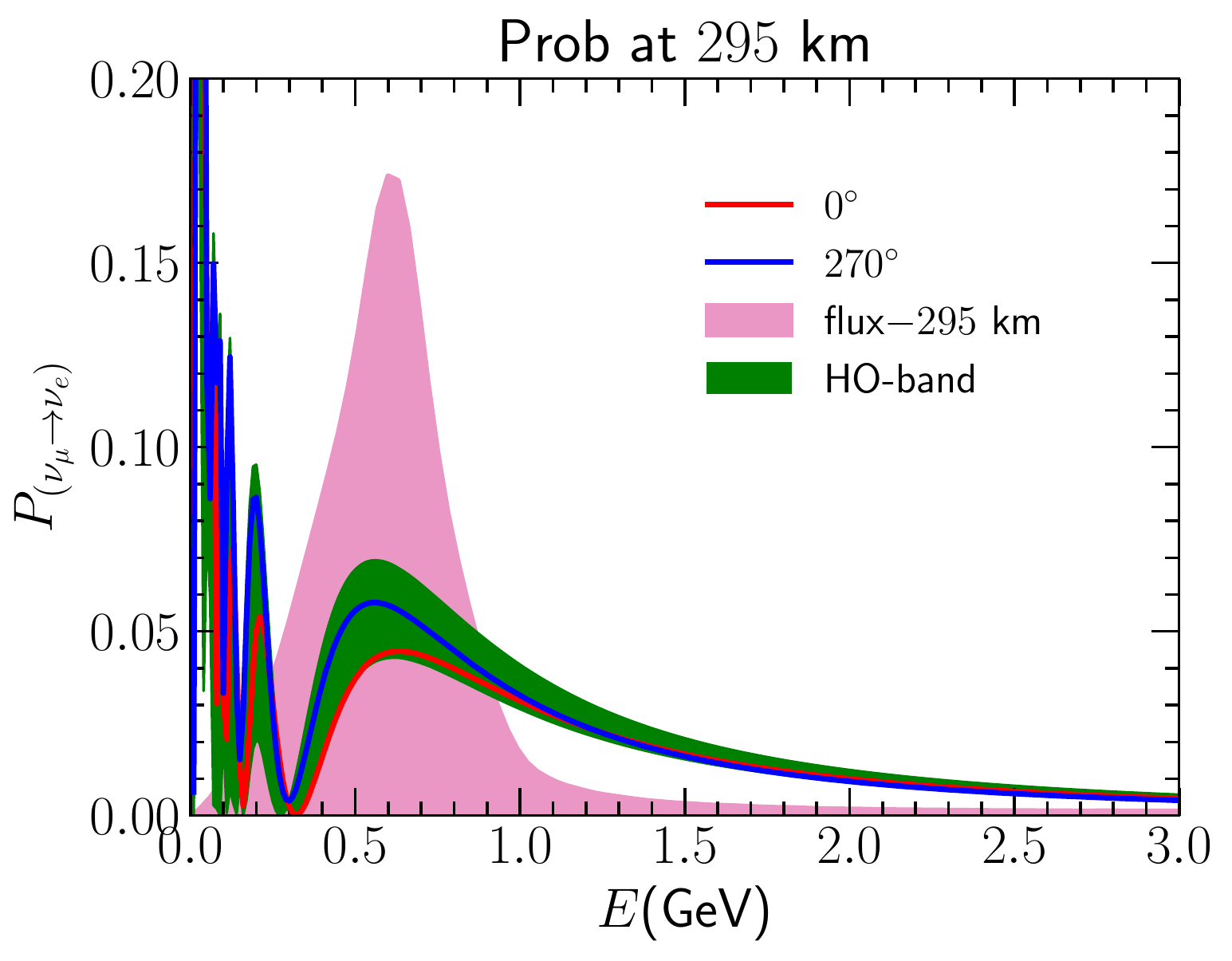}
    \hspace*{0.1 true cm}
    \includegraphics[height=55mm,width=75mm]{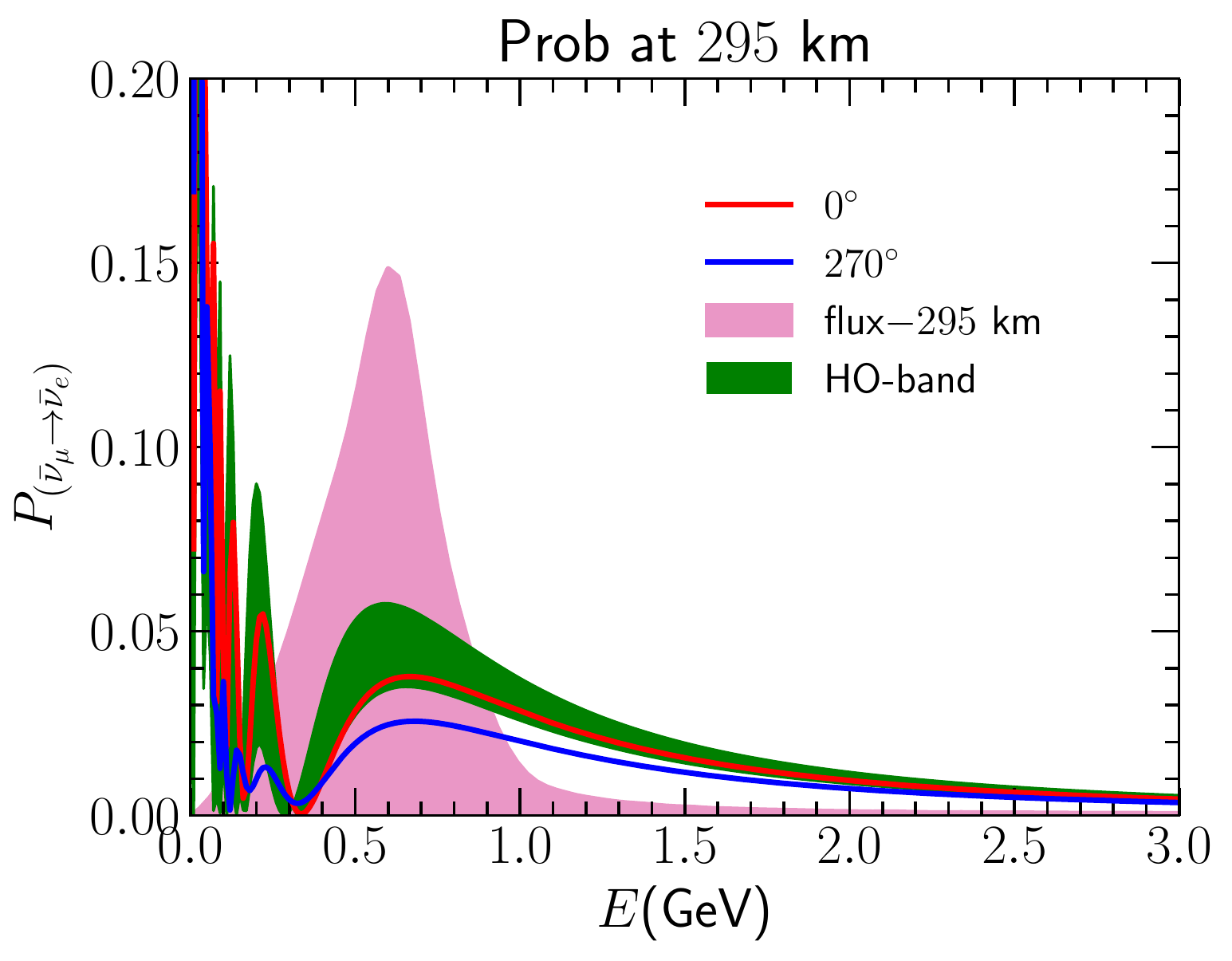} \\
    \hspace{-0.4cm}
    \includegraphics[height=55mm,width=75mm]{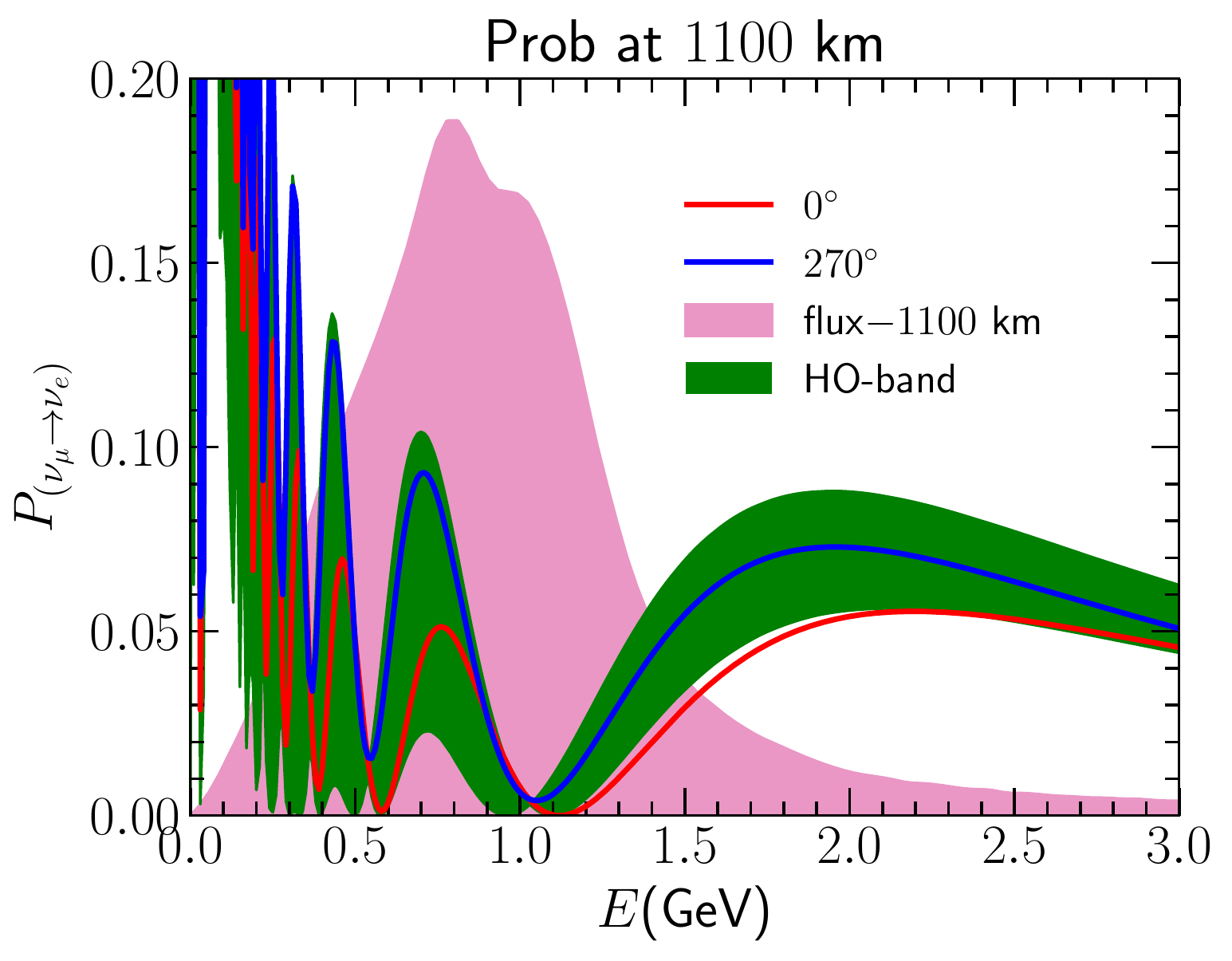}
    \hspace*{0.1 true cm}
    \includegraphics[height=55mm,width=75mm]{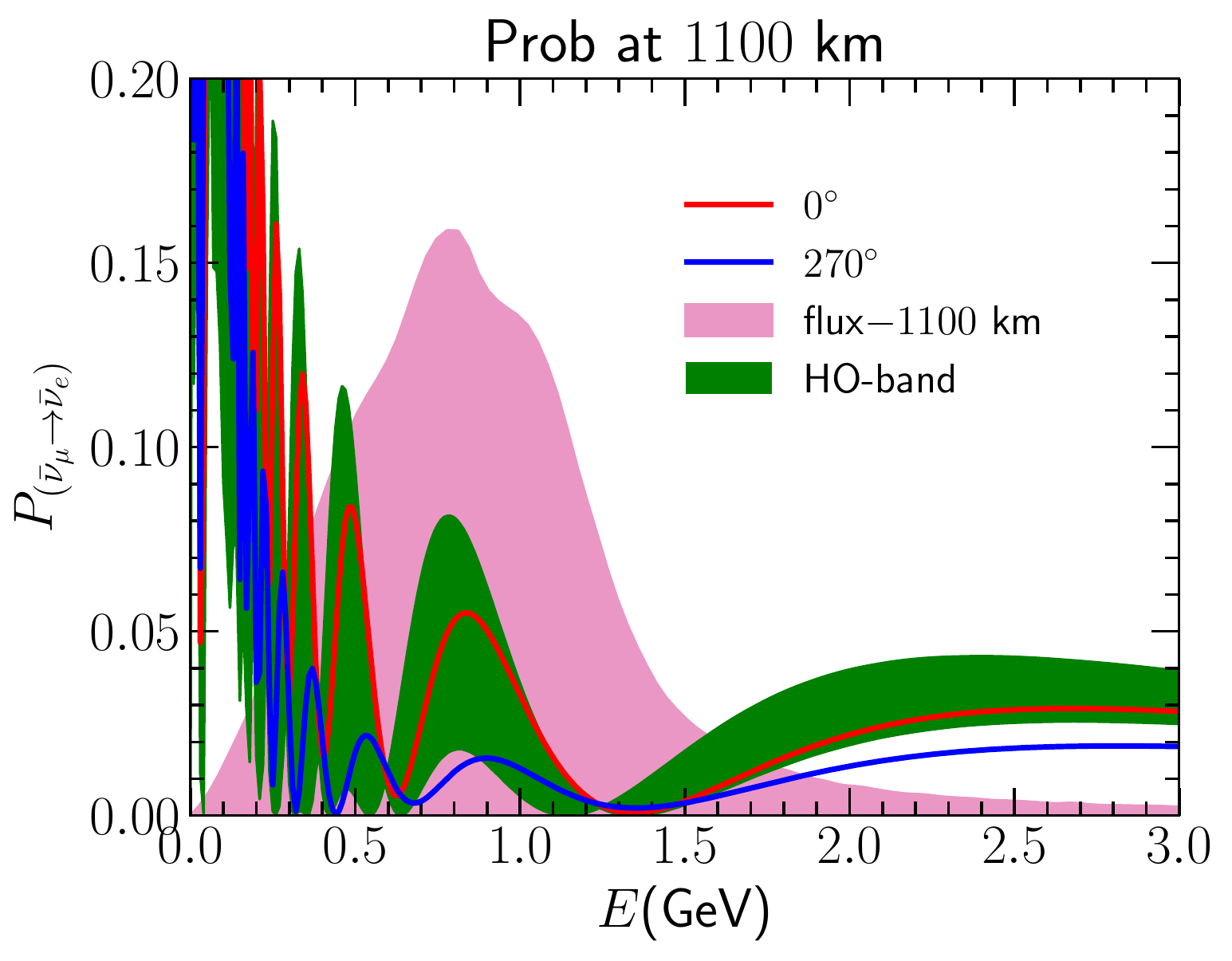}
    \caption{Upper (lower) row shows oscillation probabilities at 295 km (1100 km). Left (right) column gives the probabilities for (anti) neutrino mode. In each plot, pink shaded area represents corresponding fluxes, green shaded region is for higher octant band (HO band) and red (blue) solid line shows the probability for $\delta_{CP} = 0^{\circ} (270^{\circ})$ in lower octant.}
    \label{prob_oct}
\end{figure}

\begin{figure}[htpb]
    \centering
    \hspace{-0.4cm}
    \includegraphics[height=55mm,width=75mm]{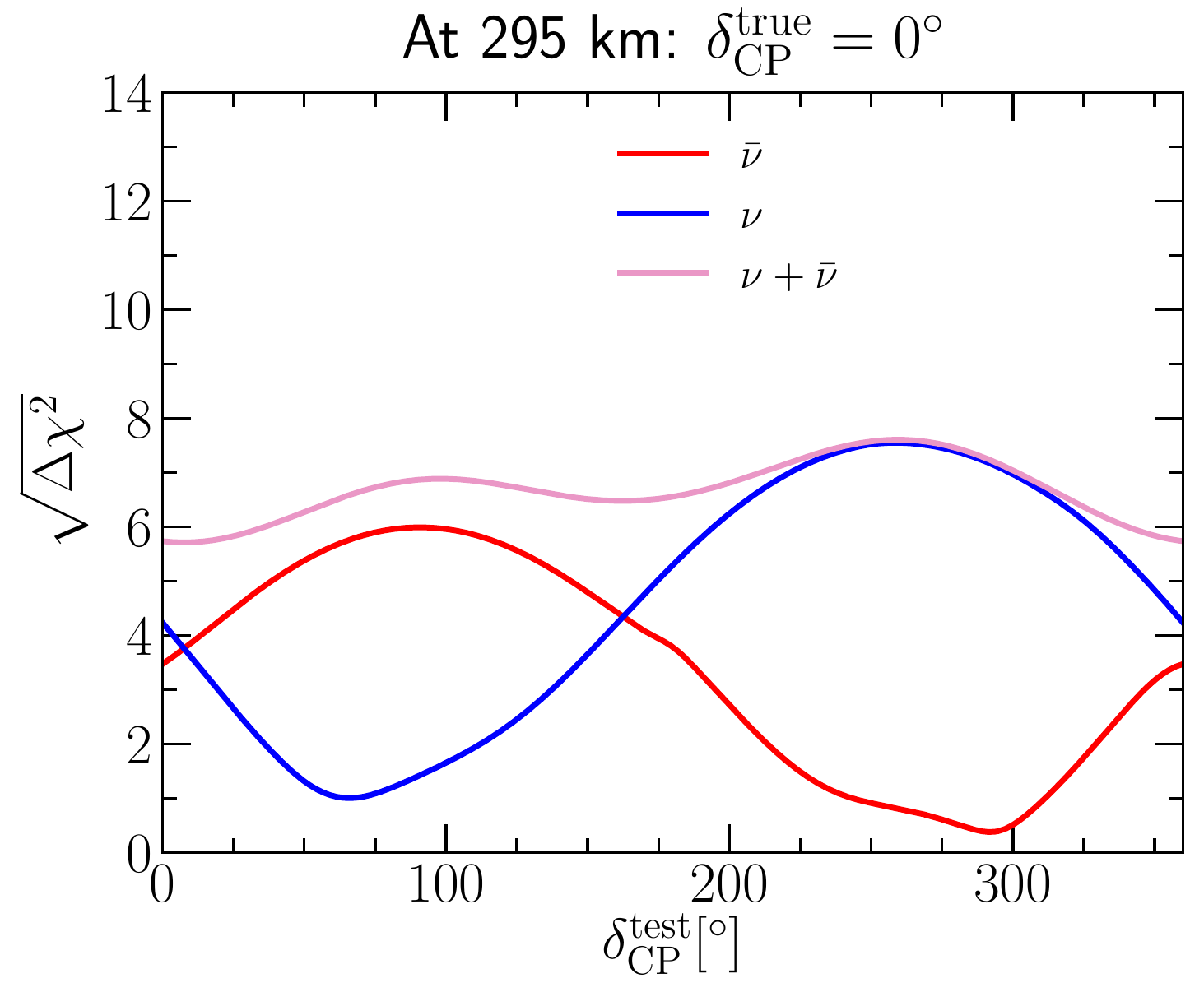}
    \hspace*{0.1 true cm}
    \includegraphics[height=55mm,width=75mm]{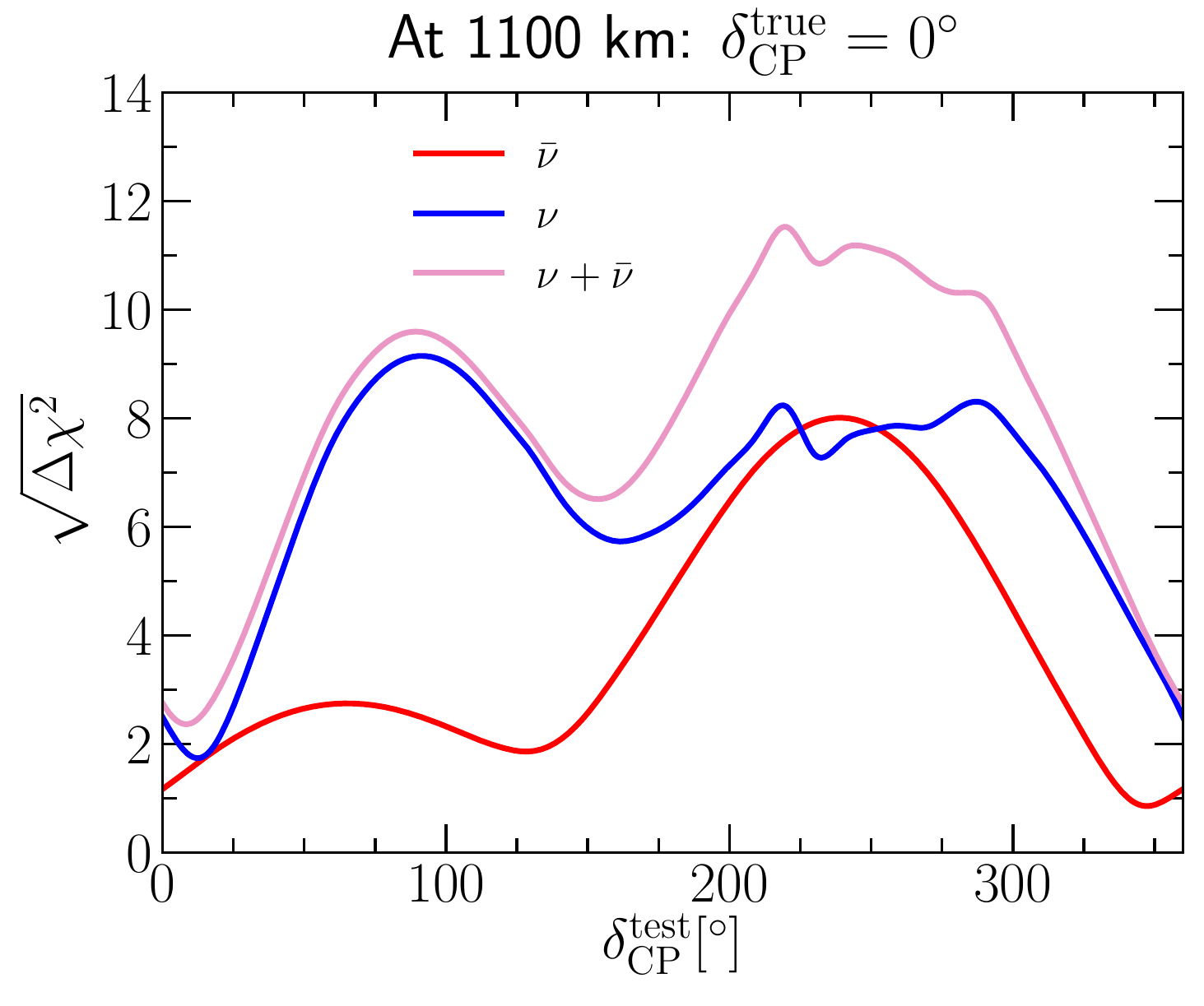}
    %\includegraphics[height=50mm,width=70mm]{theta-chi-0.pdf}
    %\hspace*{0.2 true cm}
    %\includegraphics[height=50mm,width=70mm]{theta-chi-270.pdf}
    \caption{Octant sensitivity with respect to test values of $\delta_{CP}$ in the case of hierarchy known (HK). Left (right) panel describes the variation of octant $\sqrt{\Delta \chi^2}$ when both the detectors are placed at 295 km (1100 km) with true $\delta_{\rm{CP}}=0^{\circ}$. In each plot, red (blue) [pink] solid line stands for the octant sensitivity in (anti) neutrino [neutrino + antineutrino] mode. Both the panels are generated in presence of current values of systematic errors.  }
    \label{oct_295_1100}
\end{figure}

To understand the physics of the above figure, in Fig.~\ref{prob_oct}, we have plotted the appearance channel probability as a function of energy $E$.
The top row is for 295 km and the bottom row is for 1100 km. In each row the left panel is for neutrinos and the right panel is for antineutrinos. In each panel, the pink shaded region corresponds to the flux at that baseline and polarity. In these panels, the red and blue curves are for lower octant i.e., $\theta_{23} = 42^\circ$ for two different values of $\delta_{\rm CP} = 0^\circ$ and $270^\circ$ respectively and the green band is for higher octant i.e., $\theta_{23} =$ $48^\circ$. The width of the green band is due to the variation of $\delta_{\rm CP}$ in its full range. Therefore, the separation between the red curve and the green band is proportional to the octant sensitivity for $\theta_{23} = 42^\circ$ at $\delta_{\rm CP} = 0^\circ$ and the separation between blue curve and green band is proportional to the octant sensitivity for $\theta_{23} =$ $42^{\circ}$ at $\delta_{\rm CP} = 270^\circ$. 
\begin{itemize}
    \item Let us first discuss the scenario for $\delta_{\rm CP} = 270^\circ$. For both 295 km and 1100 km, we see that the blue curve is overlapping with the green band in the neutrino probabilities but it is separated in the antineutrino probabilities. Therefore we understand that at this value of $\delta_{\rm CP}$, the octant sensitivity mainly comes from the antineutrino mode. However, the separation between the blue curve and green curve is much higher in the case of 295 km as compared to 1100 km in the energy region that is covered by the pink shaded region. That's why when we shift the detector volume to 1100 km, the sensitivity decreases (cf. upper panels of Fig.~\ref{oct_1100}). Here also we note that if the flux at 1100 km would have covered the first oscillation maximum, then the separation between the blue curve and green band would have been much higher for 1100 km as compared to 295 km.
    
    \item For $\delta_{\rm CP} = 0^\circ$, we see that the red curve and the green band are overlapping for both neutrinos and antineutrinos and for both 295 km and 1100 km. At this point, it may seem strange that though $\delta_{\rm CP} = 0^\circ$ in lower octant is degenerate with higher octant for both neutrinos and antineutrinos, the octant sensitivity is still comparable with $\delta_{\rm CP} = 270^\circ$. Actually in 295 km it so happens that the degeneracy occurs at a different values of $\delta_{\rm CP}$ for neutrinos and antineutrinos. Therefore when these channels are combined, one obtains finite octant sensitivity. However, for 1100 km, the degeneracy occurs at the same value of $\delta_{\rm CP}$ and therefore does not offer a better sensitivity than 295 km at the region covered by the flux. Therefore increasing $x$ from 0 kt, does not improve the sensitivity. This can be more clear from Fig.~\ref{oct_295_1100}.
\end{itemize}

In Fig.~\ref{oct_295_1100}, we have plotted the octant sensitivity as function of $\delta_{\rm CP}$ (test) for $\delta_{\rm CP}~(\rm true) = 0^\circ$, $\theta_{23}~(\rm true) = 42^\circ$ and for the case when the systematic errors are on. 
The left panel is for the 295 km i.e., when both the detectors are placed at Japan and the right panel is for 1100 km i.e., when both the detectors are at Korea. In these panels the blue curve is for neutrinos, the red curve is for the antineutrinos and the pink curve corresponds to the case when neutrino and antineutrinos are combined. In these panels, the $\Delta \chi^2$ minimum corresponds to the octant sensitivity at $\delta_{\rm CP} = 0^\circ$ for that particular baseline. From the left panel we see that for 295 km, the octant sensitivity of the individual neutrino and antineutrino channels are very small. For the neutrino mode, the $\Delta \chi^2$ minimum comes around $\delta_{\rm CP}$ (test) $50^\circ$ and for the antineutrino mode, the $\Delta \chi^2$ minimum comes around $\delta_{\rm CP}$ (test) around $270^\circ$. Therefore when $\Delta \chi^2$ for neutrino and antineutrino mode is added, the combined minimum $\Delta \chi^2$ occurs around $0^{\circ}$ with octant sensitivity around $6 \sigma$.  Whereas for 1100 km (right panel), we see that for both the neutrino and antineutrino mode, the $\Delta \chi^2$ minimum comes around $\delta_{\rm CP}$ (test) = $ 0^\circ$ with the value of $\sqrt{\Delta \chi^2}$ around  $3$. That is why for the combined case, the octant sensitivity remains very small.

Note that we have explained the physics of Fig.~\ref{oct_1100} for lower octant. The physics for upper octant can be explained in the similar manner.

\begin{figure}[htpb]
    \centering
    \hspace{-0.4cm}
    \includegraphics[height=55mm,width=75mm]{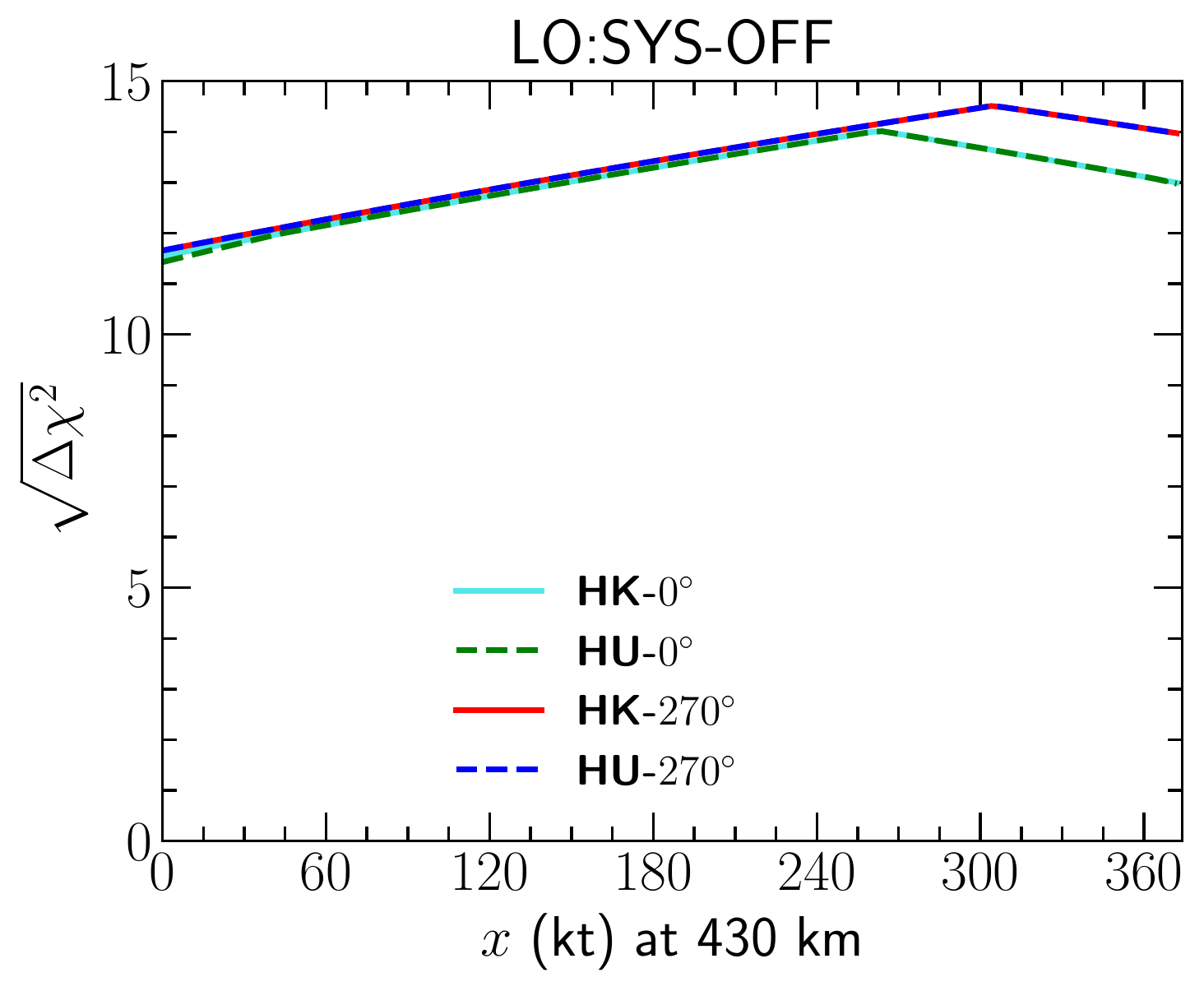}
    \hspace*{0.1 true cm}
    \includegraphics[height=55mm,width=75mm]{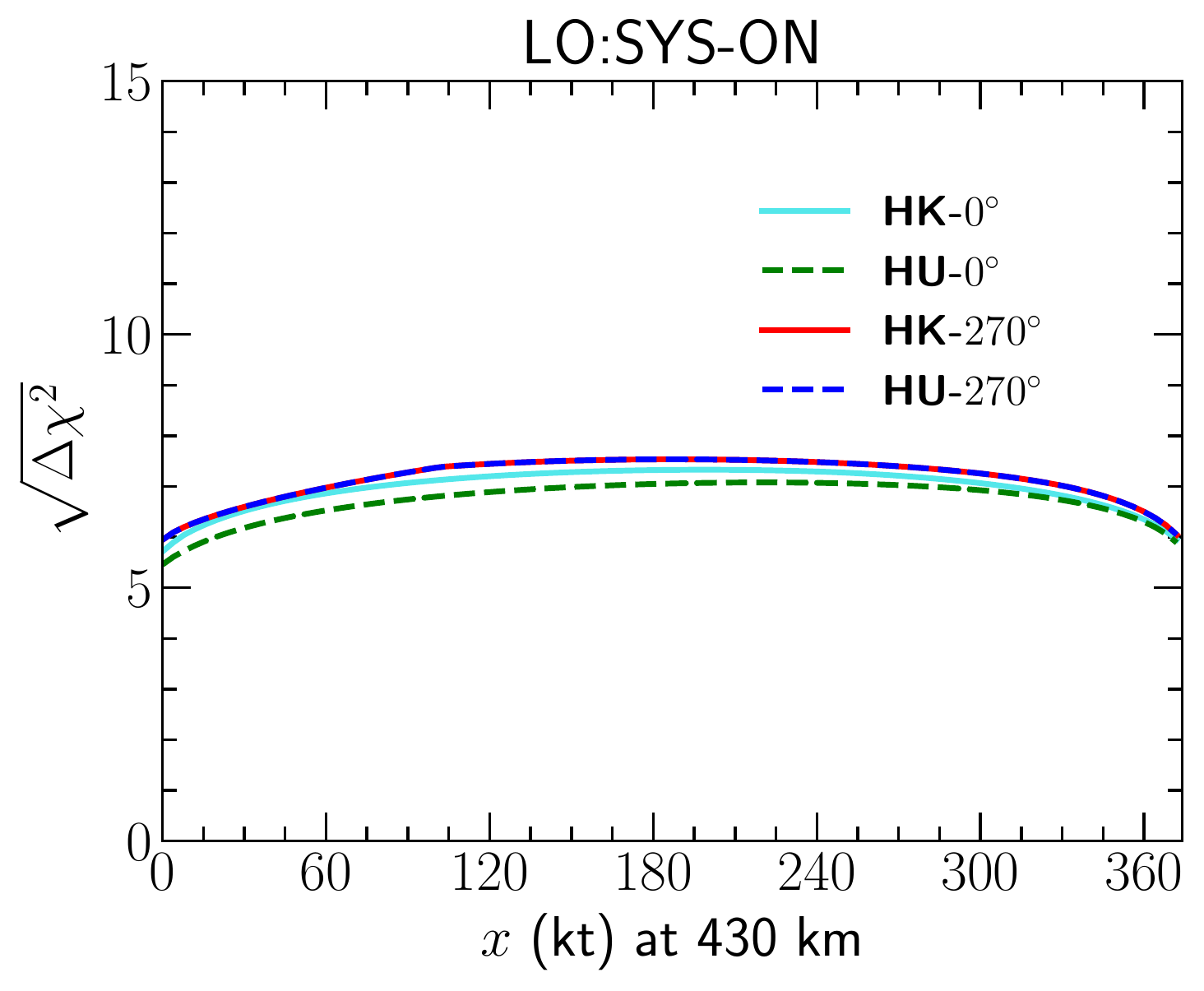} \\
    \hspace{-0.4cm}
    \includegraphics[height=55mm,width=75mm]{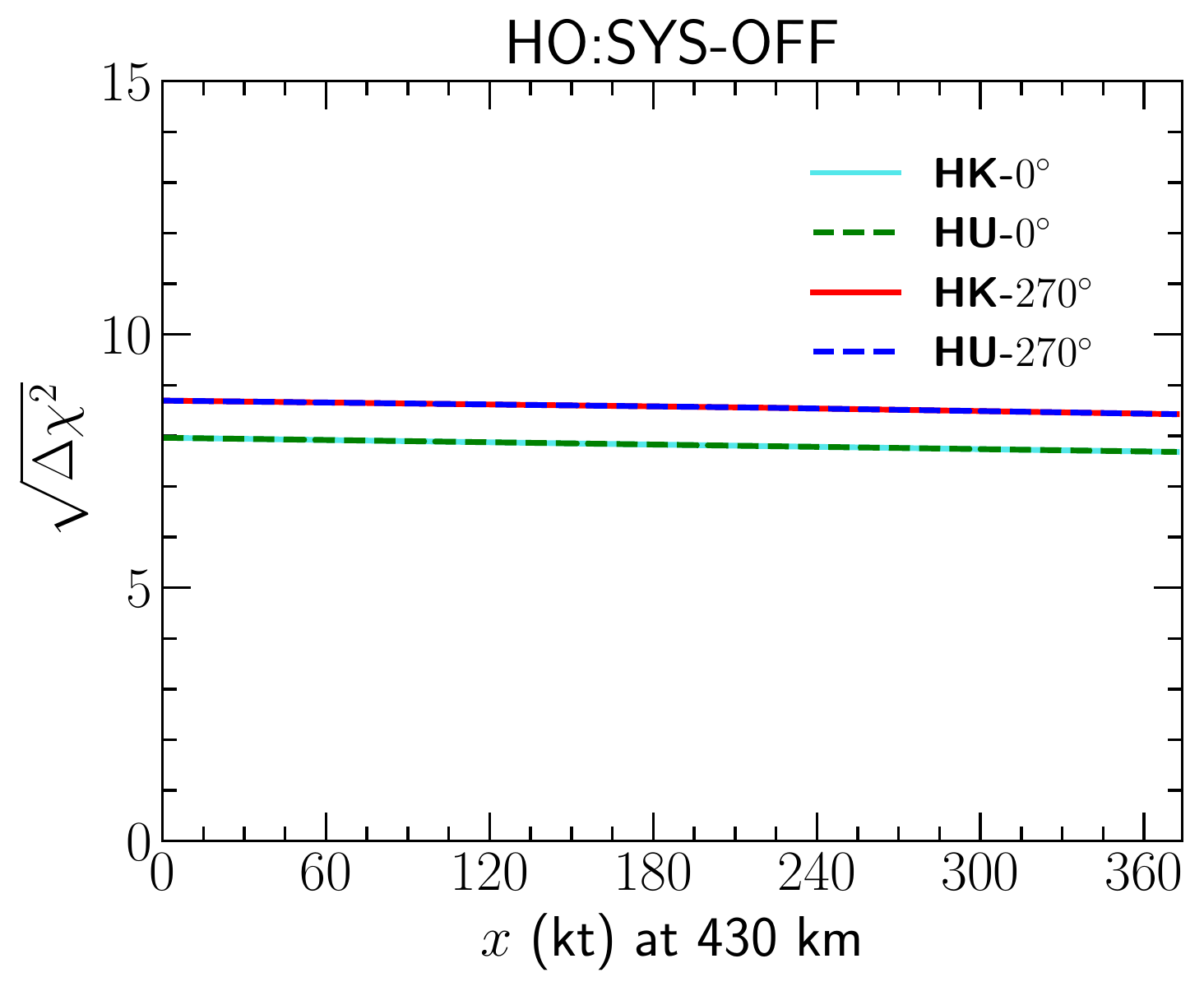}
    \hspace*{0.1 true cm}
    \includegraphics[height=55mm,width=75mm]{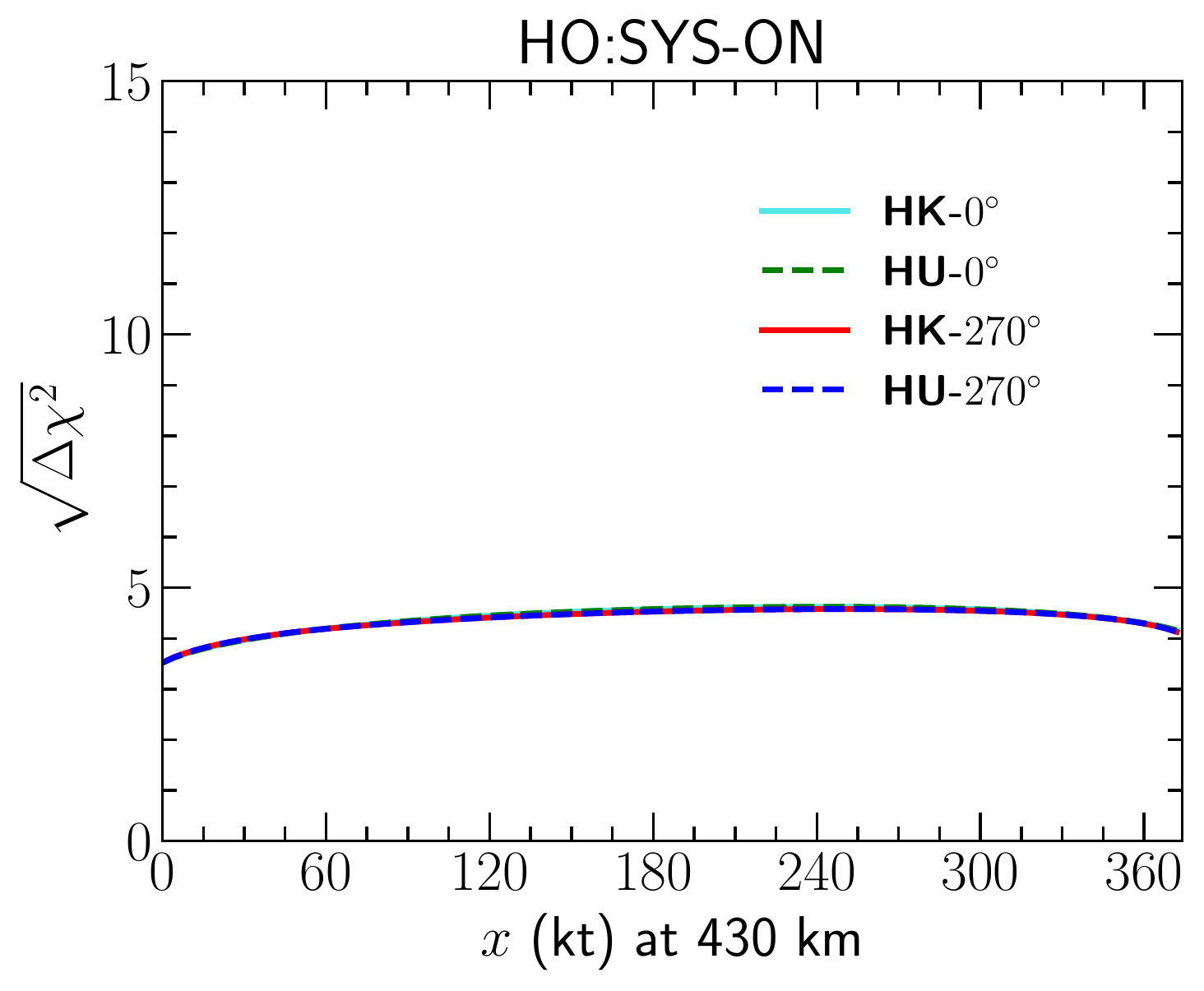}
    \caption{Octant sensitivity as a function of $x$ (detector volume at 430 km). Upper (lower) row shows octant sensitivity when $\theta_{23}^{\rm {true}}=42^{\circ} (48^{\circ})$. Left (right) column reflects the same without (with) systematic error. In each plot, `HK (HU)$-0^{\circ} [270^{\circ}]$' stands for hierarchy known (unknown) with $\delta_{\rm {CP}}^{\rm {true}} =0^{\circ} [270^{\circ}] $. }
    \label{oct_430}
\end{figure}

Now we will try to see what would have happened if the flux at 1100 km covers the first oscillation maximum in-spite of second oscillation maximum for octant. As that of hierarchy, we will consider the flux corresponding to 1100 km and use the baseline of 430 km. This we have presented in Fig.~\ref{oct_430} which is same as that of Fig.~\ref{oct_1100} except the fact that here $x$ is the detector volume at 430 km. 
In these panels we see that for lower octant, the sensitivity increases as $x$ increases from  0 kt when systematic errors are turned off (top left panel). This implies the fact that unlike the case of 1100 km, in this case additional sensitivity comes from the 430 km. When the systematic errors are turned on (top right panel), octant sensitivity increases upto a certain value of $x$ (upto $345$ kt) because of the improvement in the systematic and the additional sensitivity from the 430 km. For $x > 345 $ kt, the sensitivity starts to decrease. For higher octant, the situation is similar as that of 1100 km and this can be attributed to the fact that 430 km, does not provide any additional sensitivity for higher octant.

\subsection{CP Violation Sensitivity and CP Precision Sensitivity}

\begin{figure}[htpb]
\begin{center}
\hspace{-0.4cm}
\includegraphics[height=55mm,width=75mm]{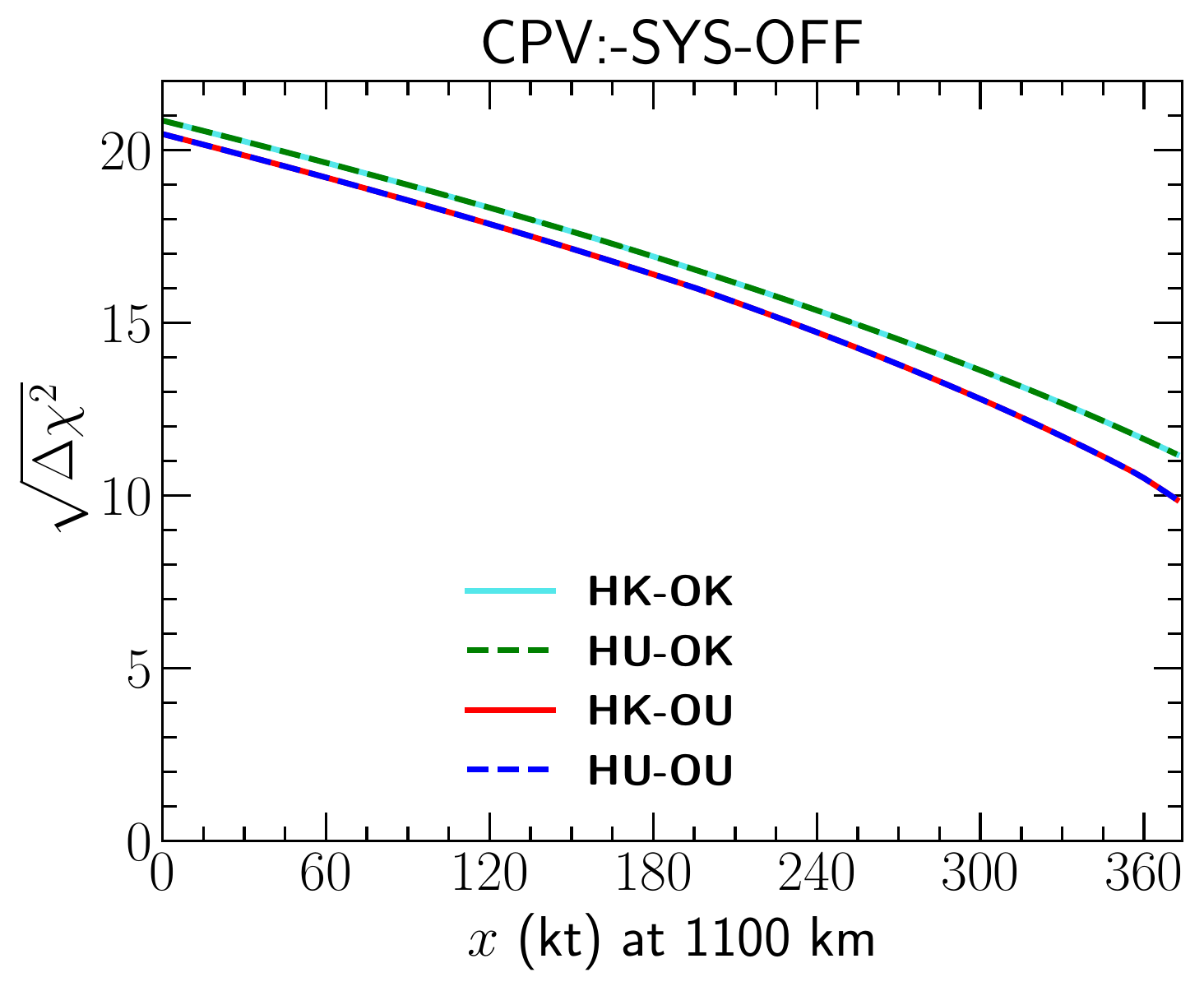}
\hspace*{0.1 true cm}
\includegraphics[height=55mm,width=75mm]{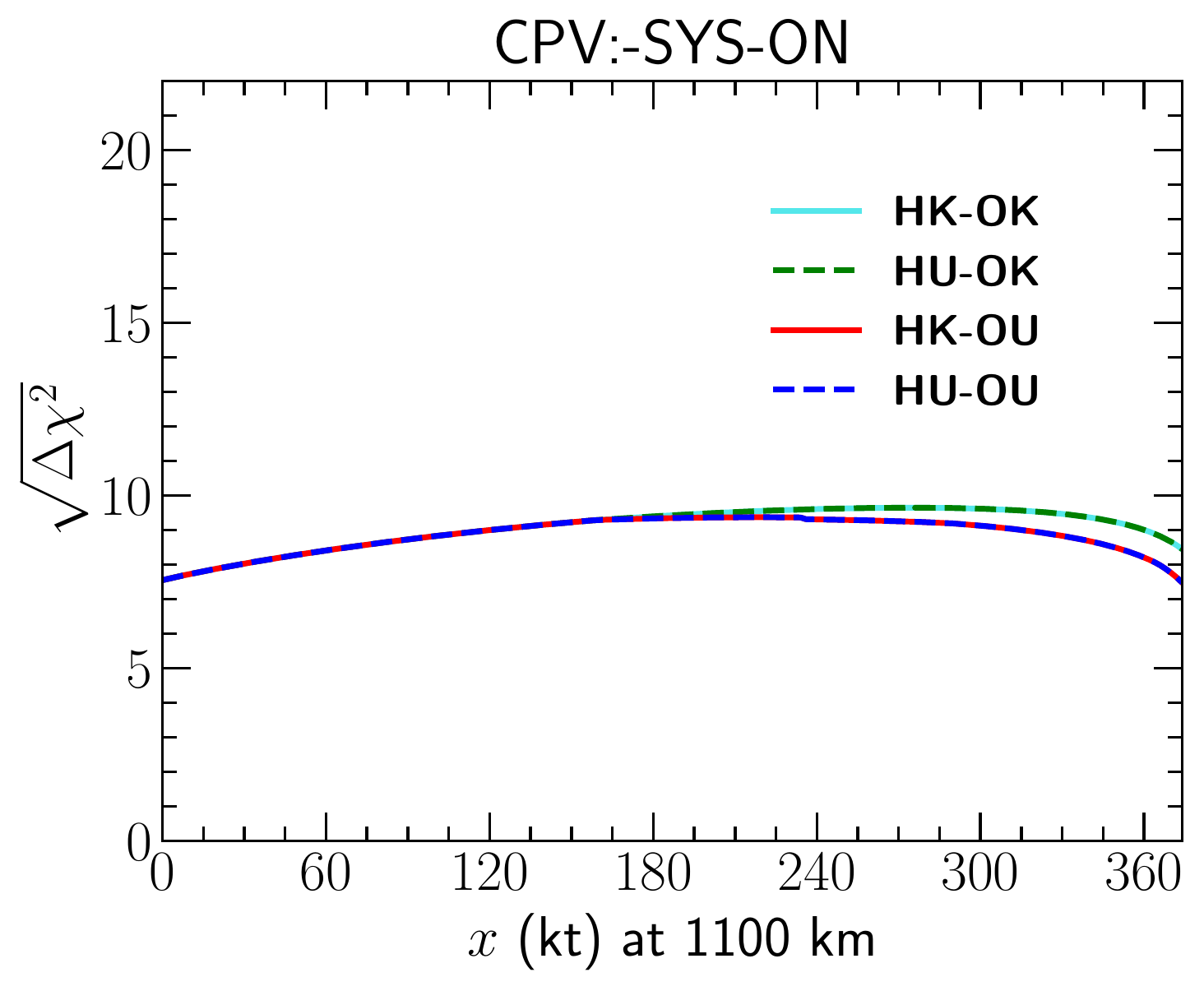}
\caption{CP violation as a function of $x$ (volume of detector at 1100 km) at $\delta_{\rm{CP}}^{\rm {true}}= 270^{\circ}$. Left (right) panel shows the CP violation variation without (with) systematic error. In each plot, `HK (HU)-OK [OU]' stands for hierarchy known (unknown) -octant known [unknown]. }
\label{cpv_1100}
\end{center}
\end{figure}

Now let us discuss the sensitivity of our setup with respect to $\delta_{\rm CP}$. CP violation discovery sensitivity is defined as the capability of an experiment to distinguish a CP violating value of $\delta_{\rm {CP}}$ from the CP conserving values $0^\circ$ and $180^\circ$. In Fig.~\ref{cpv_1100}, we have plotted the CP violation discovery sensitivity as function of $x$ where $x$ is the volume of the detector at 1100 km. 
The left panel corresponds to the case when systematic errors are switched off whereas the right panel corresponds to the case when the systematic errors are turned on. These panels are generated for true $\theta_{23} = 42^\circ$ and true $\delta_{\rm CP} = 270^\circ$. To see the effect of hierarchy degeneracy and octant degeneracy in CP violation, we have considered four cases: (i) hierarchy known - octant known (HK-OK), (ii) hierarchy known - octant unknown (HK-OU), (iii) hierarchy unknown - octant known (HU-OK) and (iv) hierarchy unknown - octant unknown (HU-OU). From this figure we see that when we switch off the systematic uncertainties, the sensitivity decreases as we increase $x$ from 0 kt. This is also a striking feature where we see that going from first maximum to second maximum does not increase the CP violation sensitivity if systematic errors are negligible. However, when the systematic errors are turned on, we see that initially the sensitivity increases as we increase $x$ from 0 kt, reaches a maximum and then it falls. Regarding degeneracy, we see that, hierarchy degeneracy does not affect the sensitivity, but octant degeneracy reduces the CP violation sensitivity to some extent. From this figure we conclude that if the systematic errors are negligible then the best CP violation sensitivity can be obtained from the T2HK setup and for the current estimated systematic uncertainties, the best CP violation sensitivity can be obtained for a volume of 255-345 kt at the 1100 km, i.e. the optimized T2HKK which is different that the original T2HKK setup. 

\begin{figure}[htpb]
    \centering
    \hspace{-0.4cm}
    \includegraphics[height=55mm,width=75mm]{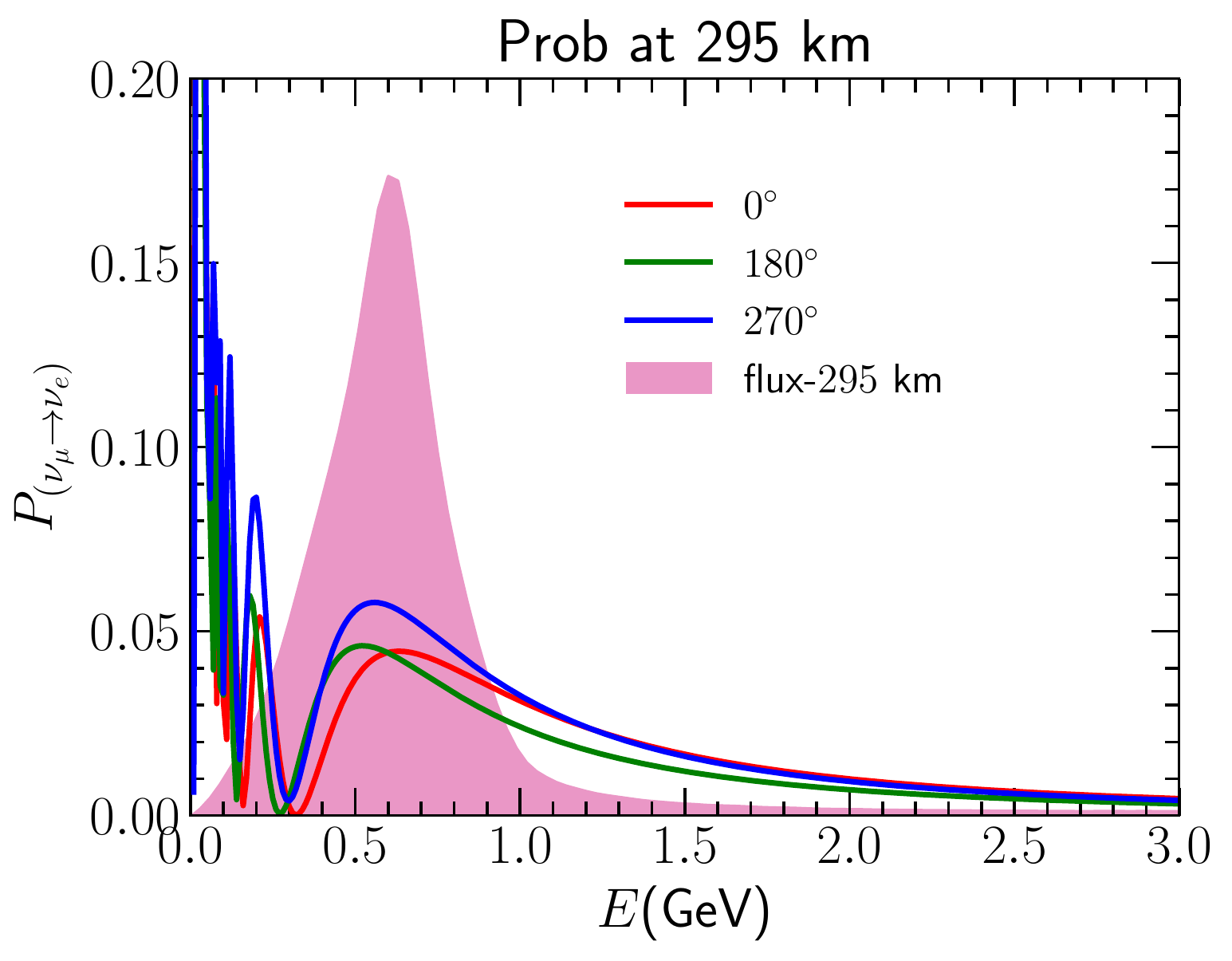}
    \hspace*{0.1 true cm}
\includegraphics[height=55mm,width=75mm]{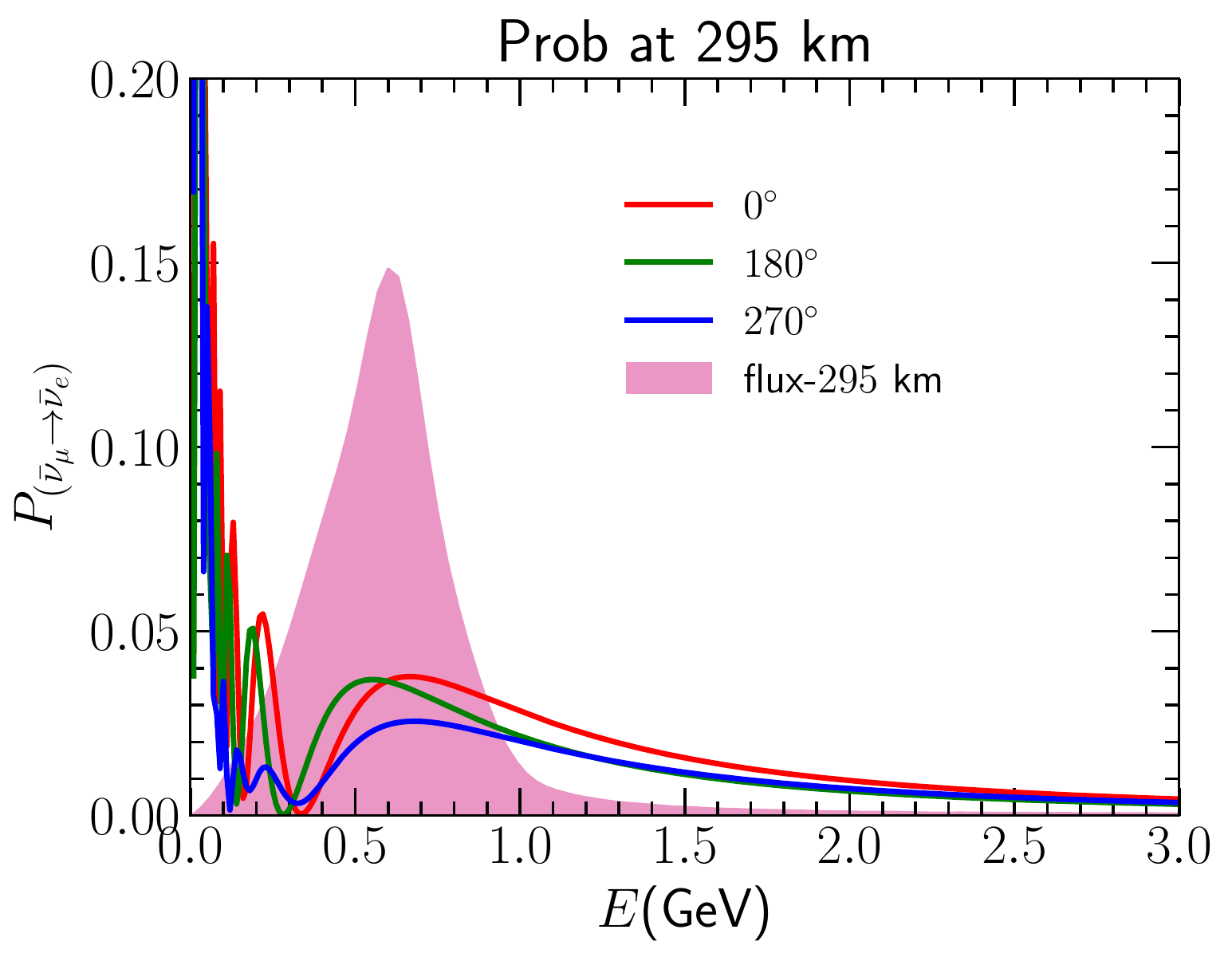}  \\
\hspace{-0.4cm}
\includegraphics[height=55mm,width=75mm]{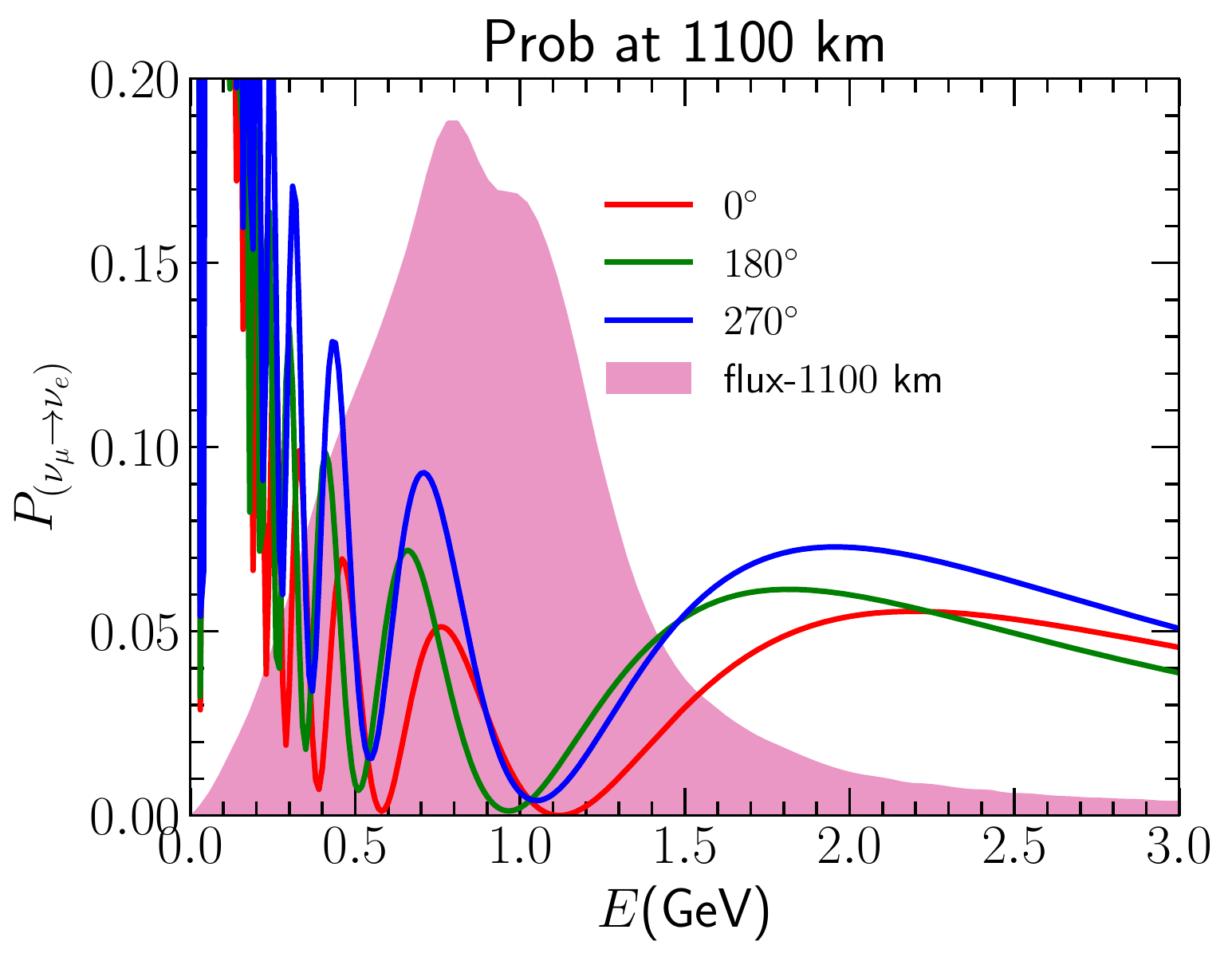}
\hspace*{0.1 true cm}
\includegraphics[height=55mm,width=75mm]{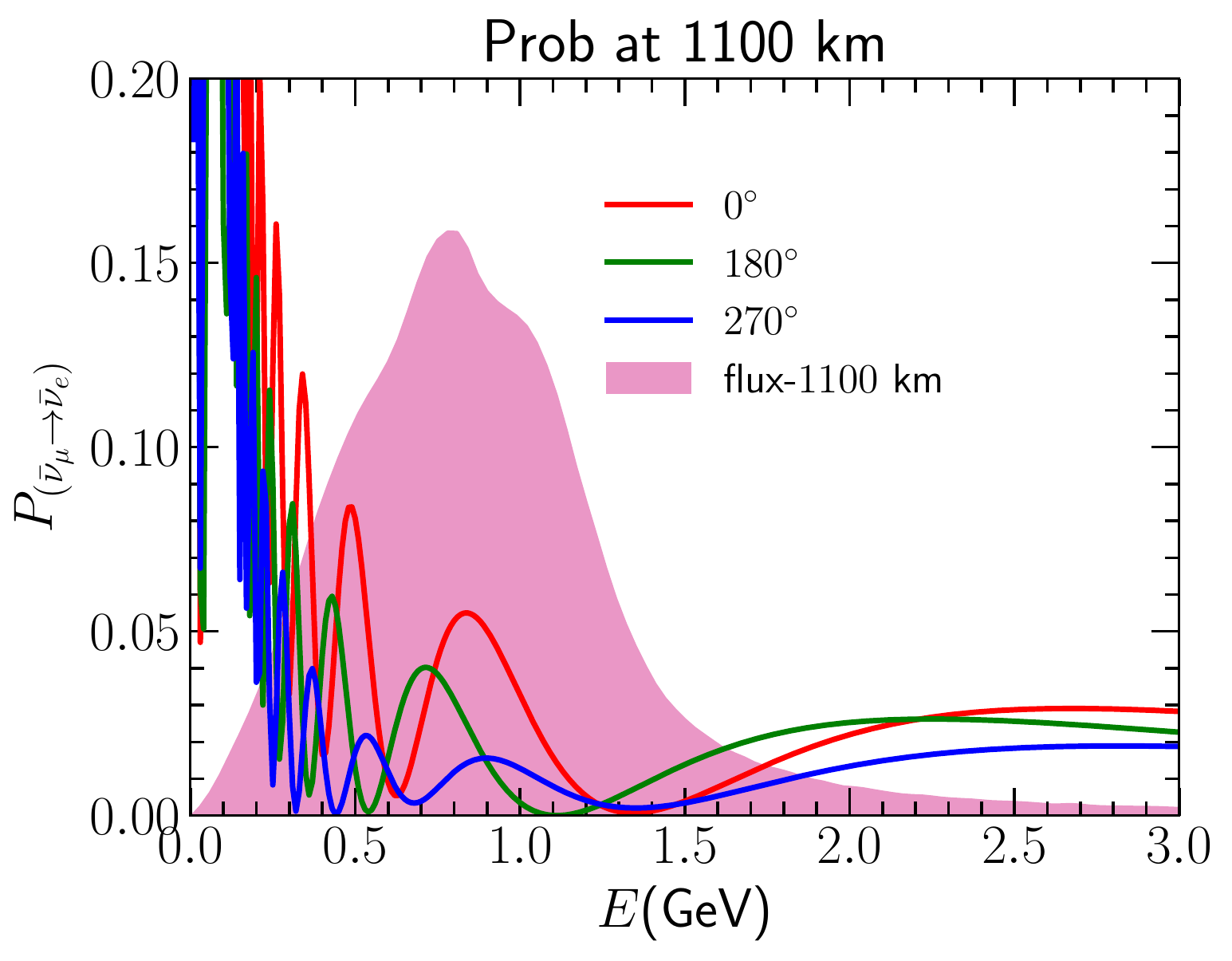}
    \caption{Upper (lower) row shows the oscillation probabilities at 295 km (1100 km) with respect to energy $E$ in GeV. Left (right) column gives the probabilities for (anti) neutrino mode. In each plot, pink shaded area represents corresponding fluxes, red, green and blue solid lines show the probability for $\delta_{\rm {CP}} = 0^{\circ}, 180^{\circ}$ and $ 270^{\circ}$ respectively.}
    \label{prob_cpv}
\end{figure}

To understand the physics behind the CP violation sensitivity at 295 km and 1100 km, in Fig.~\ref{prob_cpv}, we have plotted the appearance channel probability as a function of energy $E$ in GeV.
The top row is for 295 km and the bottom row is for 1100 km. In each row the left panel is for neutrinos and the right panel is for antineutrinos. In each panel, the pink shaded region corresponds to the flux at that baseline and polarity. In these panels, the red, green and blue curves are for $\delta_{\rm CP} = 0^\circ$, $180^\circ$ and $270^\circ$ respectively. All the panels are generated for $\theta_{23} = 42^\circ$. The separation between the blue curve and the curves corresponding to CP conserving values i.e., red and green curves are proportional to the CP violation sensitivity. From this figure we understand that the separation between the blue curve and the CP conserving curves are higher for 1100 km as compared to 295 km. This is because at the second maximum, separation between the different $\delta_{\rm CP}$ curves are higher as compared to the first oscillation maximum. But still we see that the CP violation sensitivity decreases as $x$ increases from 0 kt. This can be attributed to the fact that as we shift the detector volume from 295 km to 1100 km, the total number of events gets reduced. Hence, the improvement of the CP violation sensitivity at the second oscillation maximum cannot compensate the loss of sensitivity due to the reduction in statistics. However, when the systematic errors are turned on, it reduces the sensitivity drastically for $x = 0$ kt. This is because at this value of $x$, the sensitivity is dominated by statistics and the systematic error has a huge effect on the sensitivity. Now as we increase $x$ from 0 kt, the statistics get reduced and so is the effect of systematic errors. This causes the sensitivity to improve. Basically the drop in the sensitivity due to systematic errors is large at $x = 0$ kt and it decreases as $x$ increases. After a value of $x$ the improvement in the effect of systematic errors cannot compensate the loss of statistics and therefore the sensitivity decreases. 

\begin{figure}[htpb]
    \centering
    \hspace{-0.4cm}
    \includegraphics[height=55mm,width=75mm]{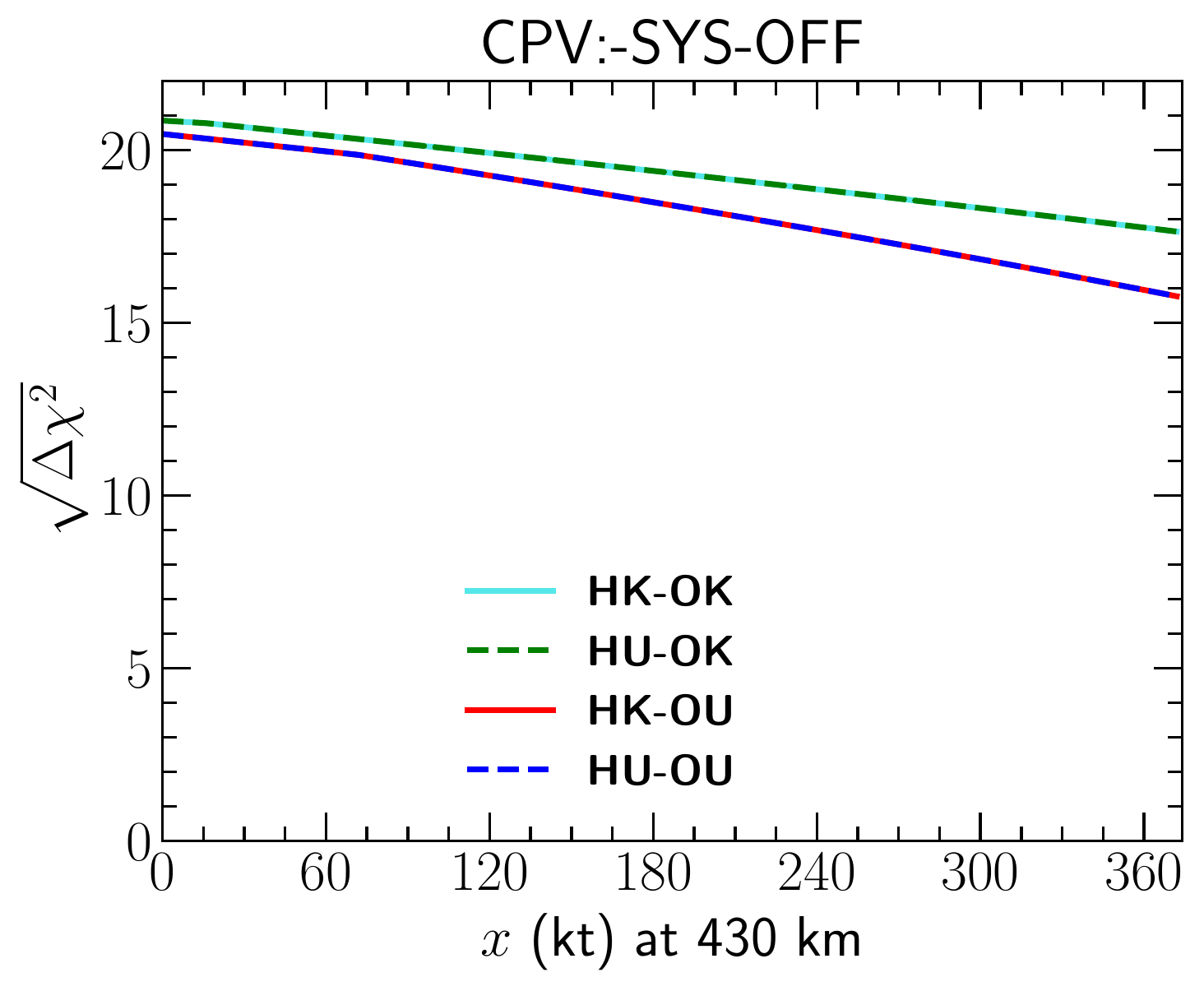}
    \hspace*{0.1 true cm}
\includegraphics[height=55mm,width=75mm]{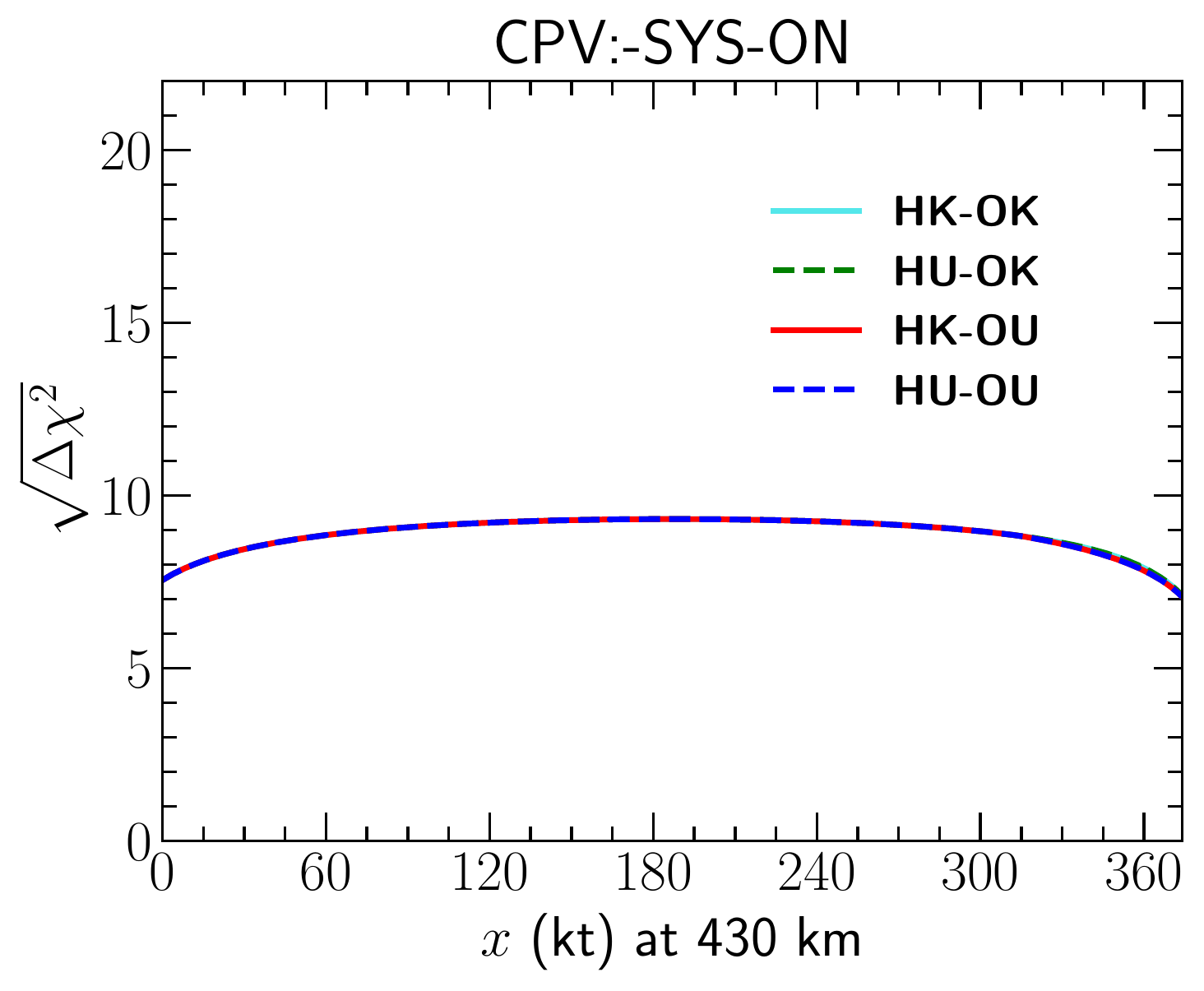}
%\includegraphics[height=50mm,width=70mm]{cpv-off-1270.pdf}
%\hspace*{0.2 true cm}
%\includegraphics[height=50mm,width=70mm]{cpv-on-1270.pdf}
    \caption{CP violation sensitivity (CPV) as a function of $x$, detector volume at 430 km. Left (right) panel plots the variation of CPV $\sqrt{\Delta \chi^2}$ for four different conditions in absence (presence) of current values of systematic errors. In each figure, `HK-OK' represents `hierarchy known-octant known', `HK-OU' stands for `hierarchy known-octant unknown', `HU-OK' depicts `hierarchy unknown-ocatnt known'  and `HU-OU' gives `hierarchy unknown-octant unknown'.}
    \label{cpv_430}
\end{figure}

Now let us try to see what would happen if we place the Korean detector at a hypothetical distance of 430 km, for which the flux of 1100 km would cover the first oscillation maximum. We have plotted this in Fig. \ref{cpv_430} which is same as that of Fig.~\ref{cpv_1100} except the fact that here $x$ is the detector volume at 430 km. 
In this figure, we see that the behaviour of the curves is same as that of Fig.~\ref{cpv_1100}. This is obvious because, the shifting of flux to first oscillation maximum from second oscillation maximum does not provide a better CP violation sensitivity which can compensate the loss of sensitivity due to the reduction of statistics when systematic errors are turned off. Also when the systematic uncertainties are turned on it gives similar effect as of Fig.~\ref{cpv_1100}.

\begin{figure}[htpb]
    \centering
    \hspace{-0.4cm}
    \includegraphics[height=55mm,width=75mm]{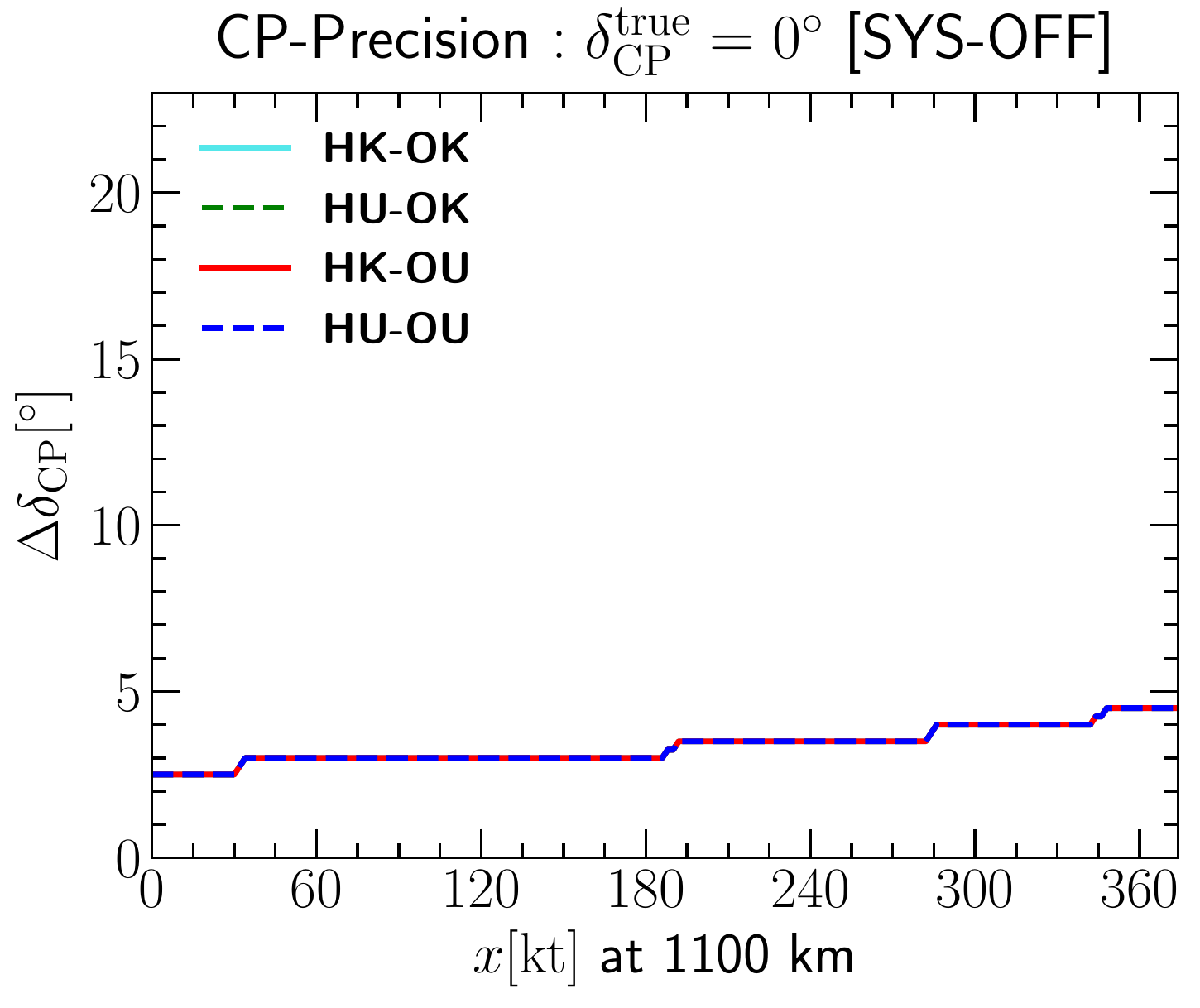}
    \hspace*{0.1 true cm}
    \includegraphics[height=55mm,width=75mm]{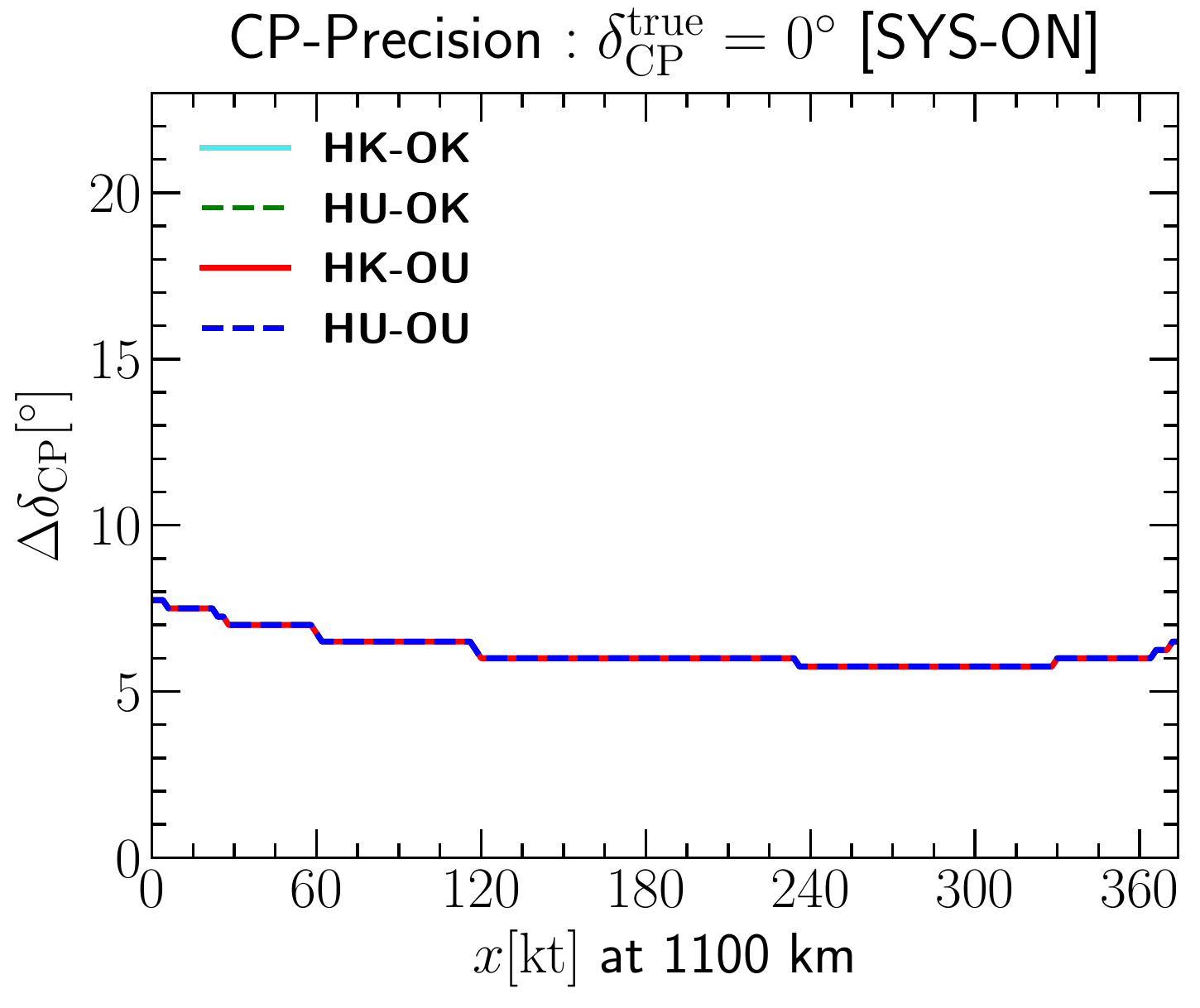} \\
    \hspace{-0.4cm}
   \includegraphics[height=55mm,width=75mm]{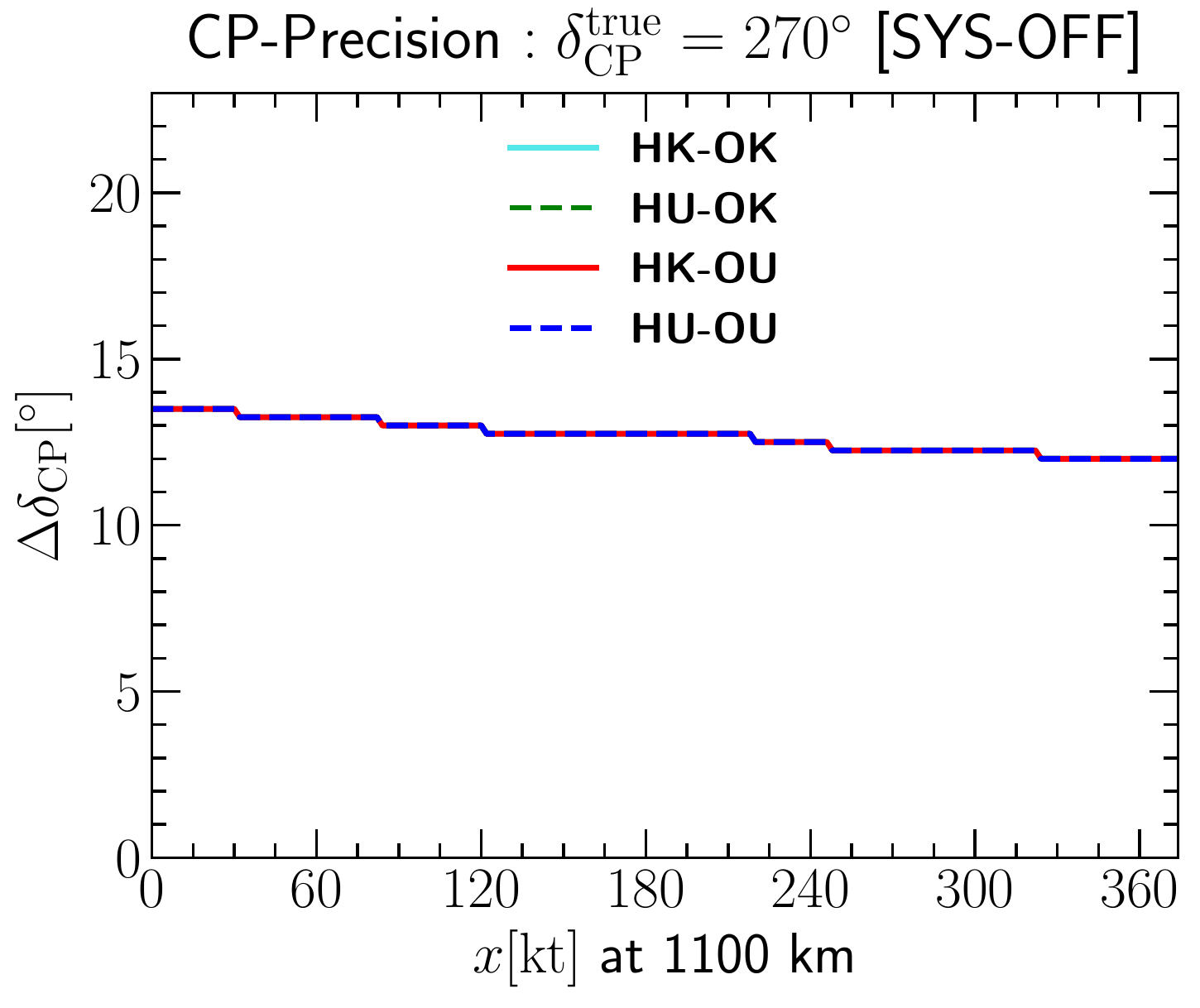}
   \hspace*{0.1 true cm}
    \includegraphics[height=55mm,width=75mm]{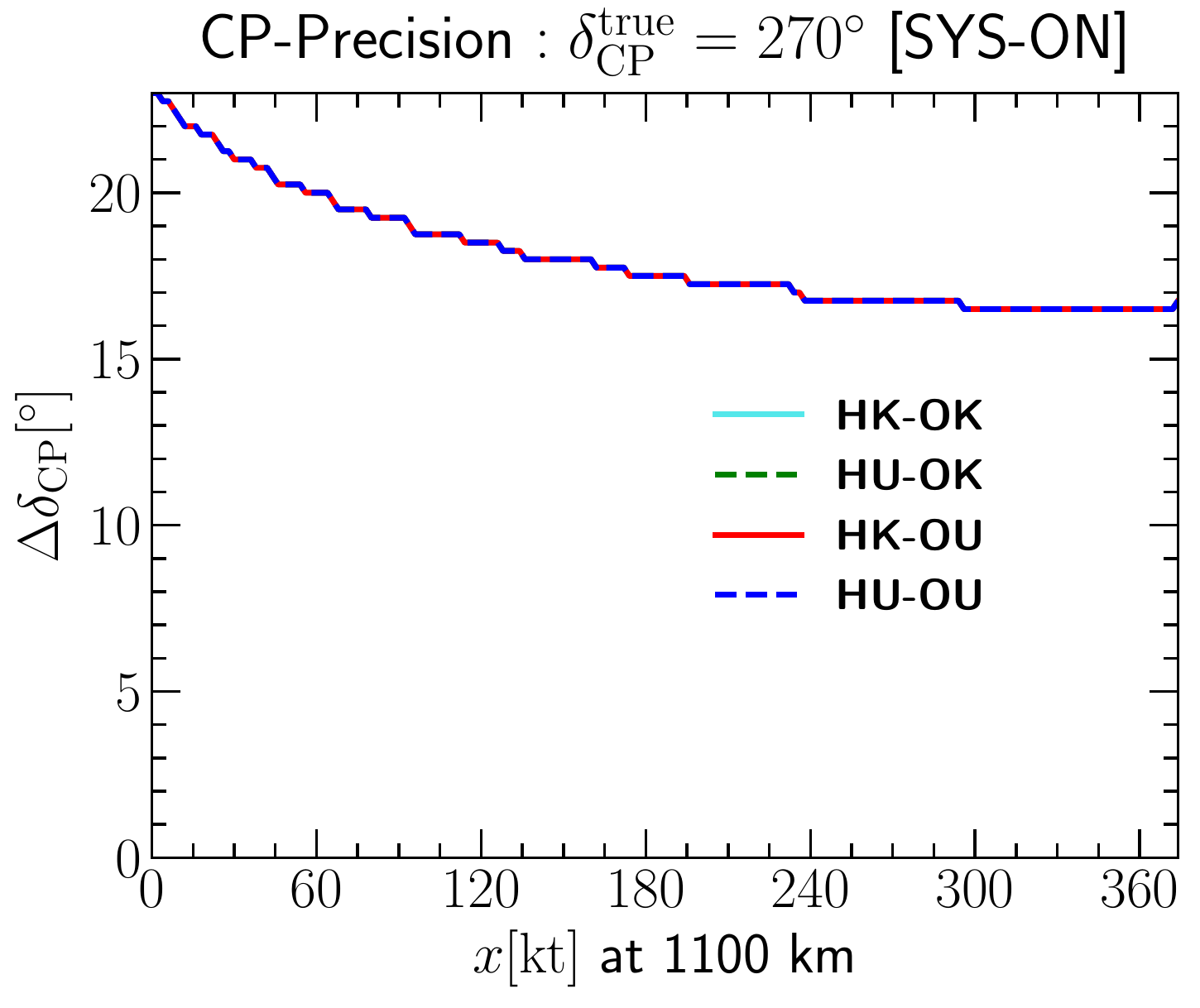}
    \caption{CP precision sensitivity as a function of $x$ (volume of the detector placed at 1100 km). Left (right) column shows the variation when systematic error is off (on). Upper (lower) panel depicts the CP precision sensitivity for $\delta_{\rm {CP}}^{\rm {true}}= 0^{\circ} (270^{\circ}).$ In each plot, `HK (HU)' represents `hierarchy- known (unknown)' and `OK (OU)' stands for `octant known (unknown)'. }
    \label{cpp_1100}
\end{figure}

Finally let us discuss the CP precision sensitivity of our setup. The CP precision sensitivity of an experiment is defined by its capability to measure a particular value of $\delta_{\rm CP}$. In Fig.~ \ref{cpp_1100}, we have plotted the half of the $ 1 \sigma$ error associated with a value of $\delta_{\rm CP}$ as a function of $x$ where $x$ is the detector volume at 1100 km. Therefore, in these panels the CP precision can be read as: $\delta_{\rm CP} \pm \Delta \delta_{\rm CP} = \delta_{\rm CP} \pm \sigma$.

In this figure, the left column is for the case when systematic errors are switched off and the right column is for the case when systematic errors are switched on. The top row corresponding to $\delta_{\rm CP} = 0^\circ$ and the bottom row corresponding to $\delta_{\rm CP} = 270^{\circ}$. For $\theta_{23}$, we have considered the true value as $42^\circ$. To see the effect of hierarchy degeneracy and octant degeneracy in CP precision, we have considered four cases: (i) hierarchy known - octant known (HK-OK), (ii) hierarchy known - octant unknown (HK-OU), (iii) hierarchy unknown - octant known (HU-OK) and (iv) hierarchy unknown - octant unknown (HU-OU). From this plot we see that when the systematic uncertainties are switched off, the sensitivity does not vary much as $x$ varies. Here we see that for $\delta_{\rm CP} = 0^\circ$, the best sensitivity can be obtained for the T2HK setup i.e., when both the detectors are placed in Japan and for $\delta_{\rm CP} = 270^\circ$ the best sensitivity can be obtained when both the detectors are in Korea. For the case when systematic errors are turned on we notice that the best sensitivity is obtained when both the detectors are placed in Korea for both values of $\delta_{\rm CP}$. Here we note that the degeneracies does not affect the CP precision at $1 \sigma$. 
 
\begin{figure}[htpb]
    \centering
    \hspace{-0.4cm}
    \includegraphics[height=55mm,width=75mm]{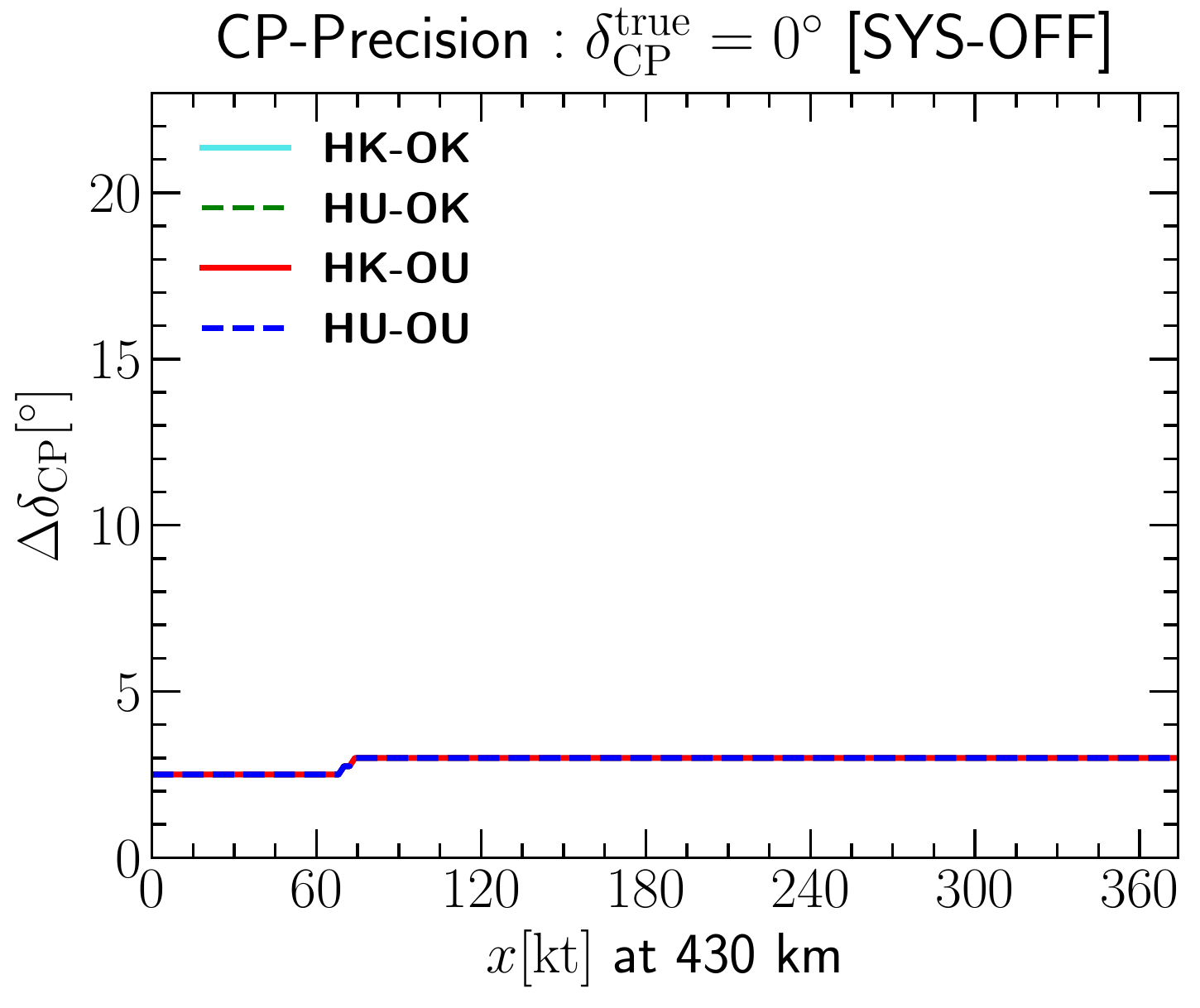}
    \hspace*{0.1 true cm}
    \includegraphics[height=55mm,width=75mm]{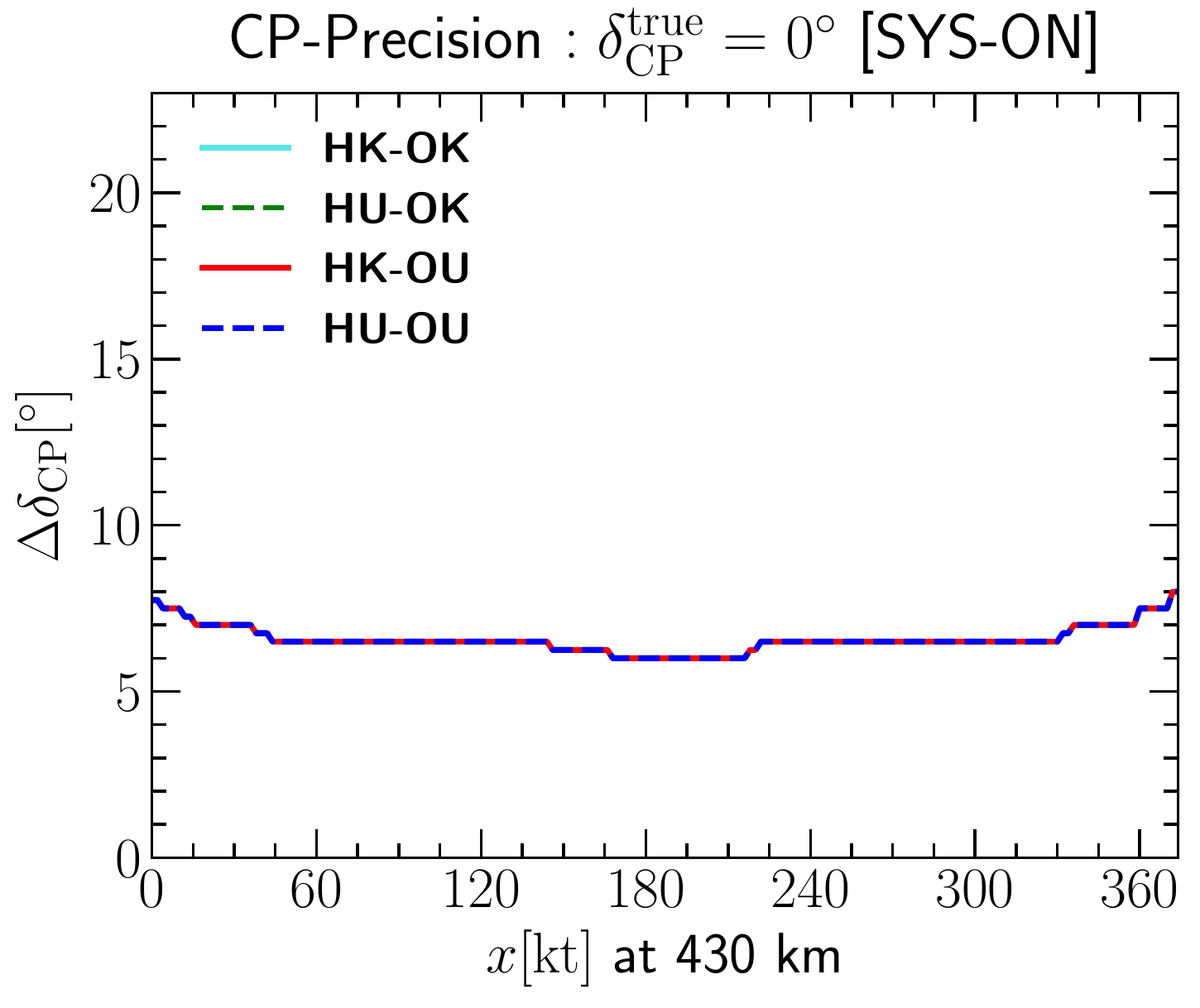} \\
    \hspace{-0.4cm}
   \includegraphics[height=55mm,width=75mm]{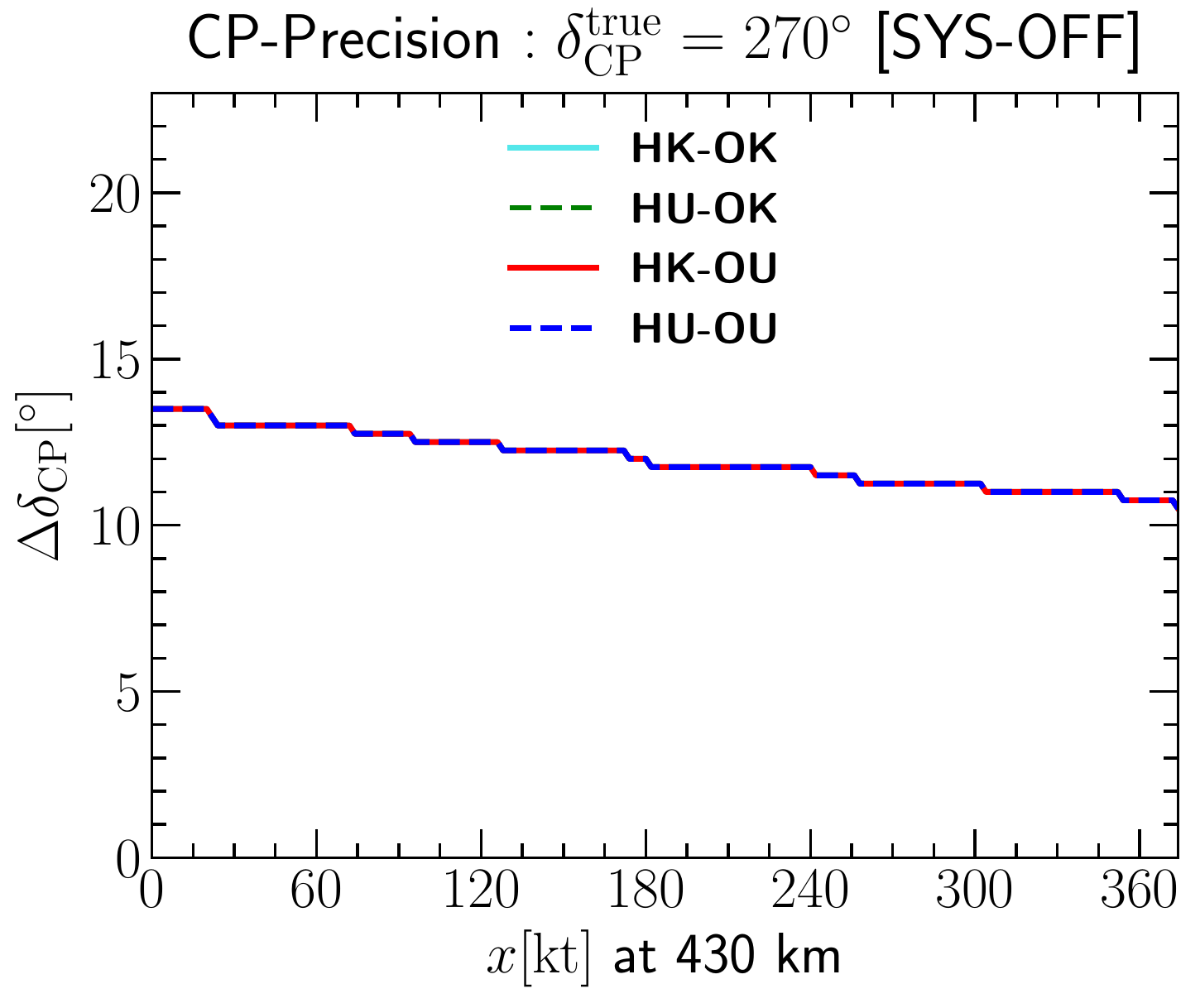}
   \hspace*{0.1 true cm}
    \includegraphics[height=55mm,width=75mm]{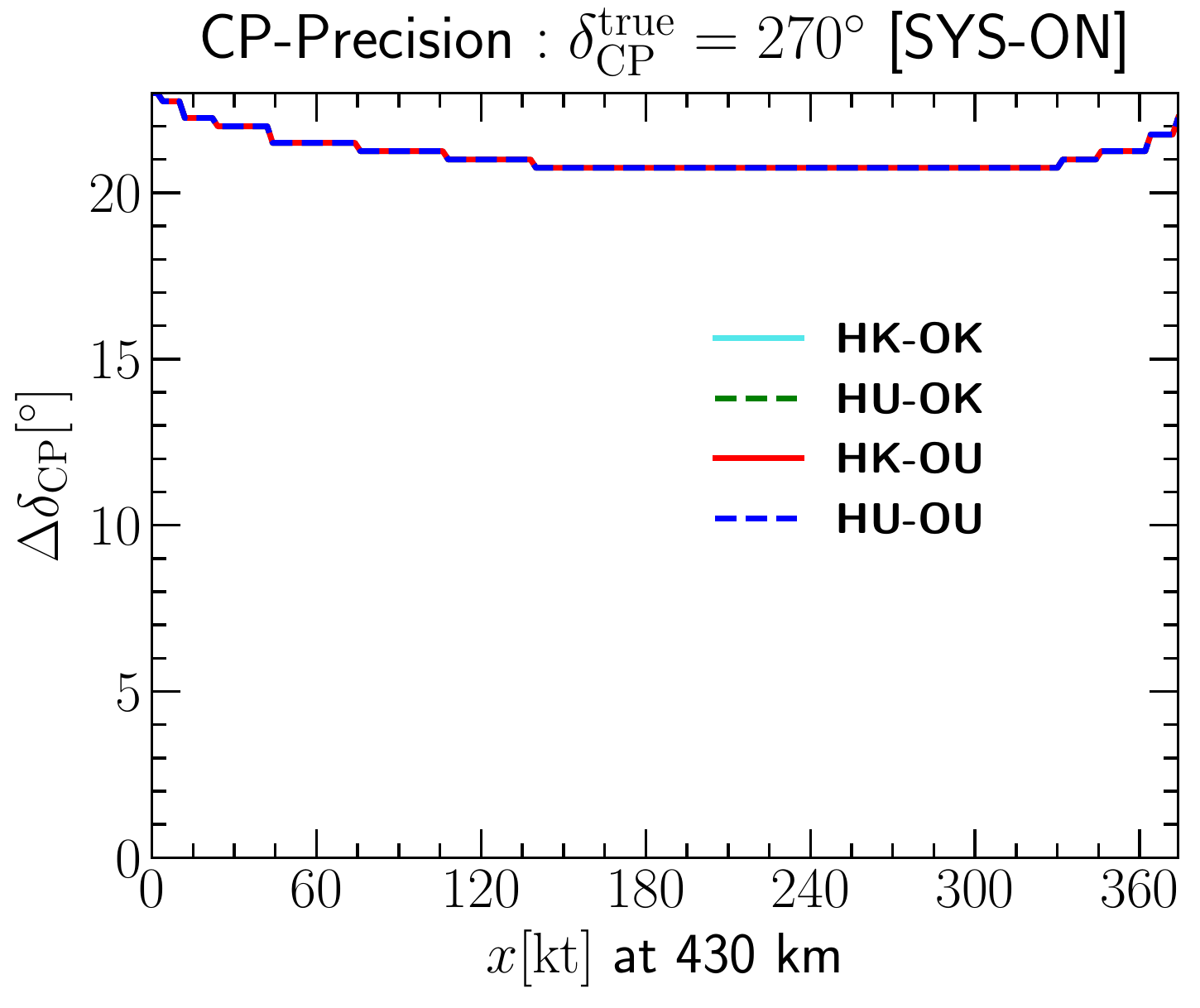}
    \caption{CP precision sensitivity as a function of $x$ (volume of the detector placed at 430 km). Left (right) column shows the variation when systematic error is off (on). Upper (lower) panel depicts the CP precision sensitivity for $\delta_{\rm {CP}}^{\rm {true}}= 0^{\circ} (270^{\circ}).$ In each plot, `HK (HU)' represents `hierarchy known (unknown)' and `OK (OU)' stands for `octant known (unknown)'.}
    \label{cpp_430}
\end{figure}

 For completeness, let us also discuss the case for 430 km by taking the flux corresponding to 1100 km. In Fig.~\ref{cpp_430}, we have plotted the same as Fig.~\ref{cpp_1100} but here $x$ corresponds to the detector volume at 430 km. 

In this figure we see almost same behaviour of the curves as that of Fig.~\ref{cpp_430}. But the sensitivity in presence of systematic uncertainties in general becomes poorer when the Koran detector is shifted to the first oscillation maximum.

{Note that curves in Figs.~\ref{cpp_1100} and \ref{cpp_430} are jagged. This is because in generating the CP precision plots, we have considered the steps of the test values of $\delta_{\rm CP}$ as $0.5^\circ$ and the steps of the detector volume as 2 kt. Therefore, the minimum change in the precision of $\delta_{\rm CP}$ is $0.25^\circ$ over a $x$-axis distance of 2 kt.}

\section{Summary and Conclusion}
\label{sum}

In this paper we discuss on the optimal detector volume of the T2HKK experiment to measure the unknowns in the standard three neutrino oscillation. T2HK is an upcoming accelerator based long-baseline experiment in Japan. Under the T2HK proposal the idea is to put two water Cherenkov detectors of 187 kt each in Japan at a distance of 295 km from the neutrino source. As an alternative option, moving a detector tank to Korea at a baseline of 1100 km is also under consideration. The flux at 295 km will cover the first oscillation maximum whereas the flux at 1100 km will cover the second oscillation maximum. The sensitivity of an experiment having dual baseline depends upon various parameters, for example: variation of statistics, effect of systematic errors, physics at second oscillation maximum and ramifications of matter density. Therefore, one needs to take these effects into consideration before choosing an optimal detector volume. In this work, we considered a variable detector volume of T2HKK considering a total detector volume of $187$ kt $\times 2  = 374$ kt and distributed it in different ratios in both the detectors to find an optimal volume for which T2HKK achieves the best sensitivity. For hierarchy and CP precision measurement, we choose the value of $\theta_{23}$ to be $42^\circ$ and choose the value of $\delta_{\rm CP}$ to be $270^\circ$ and $0^\circ$. For octant measurement we choose two values of $\theta_{23}$ which are $42^\circ$ and $48^\circ$ belonging to lower and higher octant respectively. In this case the value of $\delta_{\rm CP}$ is $270^\circ$ and $0^\circ$. For CP violation measurement we have chosen the value of $\theta_{23}$ as $42^\circ$ and $\delta_{\rm CP}$ as $270^\circ$. 

In our analysis we find that the optimal volume of T2HK depends upon the value of systematic error, true value of $\delta_{\rm {CP}}$, flux at the 1100 km and the sensitivity under consideration. For hierarchy, we find that if the systematic errors are reduced to negligibly small number then the T2HK setup i.e., when both the detectors are placed at a distance of 295 km, gives best sensitivity for $\delta_{\rm CP}$ is $270^\circ$ unless the flux of the 1100 km is optimized for the first oscillation maximum. However, for $\delta_{\rm CP}$ is $0^\circ$ and if the systematic errors remain around the current values then the best sensitivity can be obtained if the detector volume is around 255-345 kt. For octant, we find that the best sensitivity comes from the T2HK setup irrespective of the value of  systematic errors unless the flux of the 1100 km is optimized for the first oscillation maximum. For the case where we can have the effect of first oscillation maximum with 1100 km flux (equivalent to a hypothetical distance of 430 km from the source J-PARC), one have better sensitivity for lower octant if the detector volume at 430 km is around 255-345 kt. For CP violation, we find that if the systematic errors are negligible then the best CP violation sensitivity can be obtained from the T2HK setup and for the current estimated systematic uncertainties, the best CP violation sensitivity can be obtained for a volume of 255-345 kt at 1100 km. For CP precision, we find that best sensitivity obtained when both the detectors are placed in Korea except for $\delta_{\rm CP} = 0^\circ$ and systematic errors are switched off. For that particular case best sensitivity will be obtained when both the detectors are placed in Japan. We have summarized our findings in Table~\ref{table_sum} corresponding to the current baselines and current flux options of T2HKK experiment.  
\begin{table}[htpb]
\hspace{-0.7cm}  
\begin{tabular}{|l|l|l|l|l|l|l|l|l|}
    \hline
    \multirow{2}{*}{Setup} &
      \multicolumn{2}{c|}{Hierarchy} &
      \multicolumn{2}{c|}{Octant(LO, HO)} &
      \multicolumn{2}{c|}{CP violation}  &
      \multicolumn{2}{c|}{CP precision}    \\
    & Sys off & Sys on & Sys off & Sys on & Sys off & Sys on & Sys off & Sys on \\
    & $0^\circ$ $(270^\circ)$  & $0^\circ$ $(270^\circ)$  &  $0^\circ$ $(270^\circ)$ & $0^\circ$ $(270^\circ)$  &  $270^\circ$  &  $270^\circ$  &  $0^\circ$ $(270^\circ)$  &   $0^\circ$ $(270^\circ)$   \\
    \hline
    T2HK & $\times$ ($\checkmark$) & $\times$($\times$) & $\checkmark$($\checkmark$) & $\checkmark$ ($\checkmark$) & $\checkmark$ & $\times$  & $\checkmark$($\times$) & $\times$($\times$) \\
    \hline
    T2HKK & $\times$($\times$) & $\times$($\times$) & $\times$($\times$) & $\times$($\times$) & $\times$ & $\times$ &$\times$($\times$) & $\times$($\times$) \\
    \hline
    Optimal & $\checkmark$($\times$) & $\checkmark$ ($\checkmark$) & $\times$($\times$) & $\times$($\times$) & $\times$ & $\checkmark$  &  $\times$($\checkmark$)  &  $\checkmark$($\checkmark$) \\
     T2HKK &  &  &  &  & &   &   & \\
    \hline
  \end{tabular}
  \caption{Optimal detector configuration for hierarchy, octant and $\delta_{\rm CP}$ sensitivity. Optimal (not optimal) configuration is expressed as $\checkmark (\times)$ sign. T2HK setup implies when two detectors with 187 kt each are placed at 295 km. T2HKK setup implies when one of the detector with 187 kt is at shifted to Korea. The optimal T2HKK is the setup where one tank with the range of (29-119) kt volume placed at 295 km and the other detector with the range of (255-345) kt volume placed at 1100 km from the source.}
  \label{table_sum}
\end{table}

In conclusion we want to state that finding an optimal volume for the T2HKK setup is a non trivial task and simply putting 187 kt in Japan and 187 kt in Korea does not provide the best sensitivity. Depending upon the true value of $\delta_{\rm CP}$,  values of systematic errors and the unknown parameters under consideration, one needs to decide the detector volume at Korea.

\section*{Acknowledgements}

PP and PM want to acknowledge Prime Minister's Research Fellows (PMRF) scheme for financial support. MG acknowledges Ramanujan Fellowship of SERB, Govt. of India, through grant no: RJF/2020/000082. RM would like to acknowledge University of Hyderabad IoE project grant no: RC1-20-012. We gratefully acknowledge the use of CMSD HPC facility of Univ. of Hyderabad to carry out the computational works. We also acknowledge Dinesh Kumar Singha for useful discussions.

\bibliographystyle{JHEP}
\bibliography{optimal_T2HK}

\end{document}